%% file: Arxiv_Network_deniability_v7.tex
\documentclass[journal,onecolumn,12pt,twoside]{IEEEtranTCOM}
\normalsize

\ifCLASSINFOpdf
  \usepackage[pdftex]{graphicx}
  \graphicspath{{/Drawings/}}
\else
  \usepackage[dvips]{graphicx}
  \graphicspath{{/Drawings/}}
\fi

\usepackage[tight,footnotesize]{subfigure}
\usepackage{amsmath, amsfonts, amssymb, amscd, amsthm, xspace}

\usepackage{color}
\usepackage{graphicx}
\usepackage{epstopdf}

\newcommand{\argmax}{\operatornamewithlimits{argmax}}

\newtheorem{definition}{Definition}
\newtheorem{theorem}{Theorem}

\newtheorem{lemma}{Lemma}

\theoremstyle{remark}
\newtheorem{remark}{Remark}

\newcommand{\Oh}[1]{\mathcal{O}\left(#1 \right)}
\newcommand{\oh}[1]{{o}\left(#1 \right)}

\newcommand{\Hp}[1]{H\left( #1 \right)}
\newcommand{\Hcond}[2]{H\left( {#1} | {#2} \right)}
\newcommand{\I}[2]{I\left( {#1} ; {#2} \right)}
\newcommand{\D}[2]{{D}\left({#1} || {#2}\right)}
\newcommand{\Vdist}[2]{\mathbb{V}\left({#1}, {#2}\right)}

\newcommand{\R}{R} 
\newcommand{\Rs}{R_s} 
\newcommand{\Rd}{R_d} 
\newcommand{\Rhd}{R_{h,d}} 

\newcommand{\Capd}{C_{d}} 
\newcommand{\Caphd}{C_{h,d}} 

\newcommand{\C}{C} 
\newcommand{\W}{\mathcal{W}} 
\newcommand{\Wc}{\mathcal{W}^c} 
\newcommand{\bigW}{\mathfrak{W}} 

\newcommand{\enc}{Enc} 
\newcommand{\code}{\mathcal{C}} 
\newcommand{\T}{\mathbf{T}} 

\newcommand{\codeW}{\code_{\vxW}}

\newcommand{\TW}{\hat{\T}_W} 
\newcommand{\TB}{\hat{\T}_B} 
\newcommand{\vMW}{\hat{\mathbf{M}}_W} 
\newcommand{\vMB}{\hat{\mathbf{M}}_B} 

\newcommand{\M}{M}
\newcommand{\m}{m}
\newcommand{\vM}{\vec{\mathbf{M}}} 
\newcommand{\vm}{\vec{\mathbf{m}}} 
\newcommand{\vX}{\vec{\mathbf{X}}} 
\newcommand{\vx}{\vec{\mathbf{x}}} 
\newcommand{\vxdash}{\vec{\mathbf{x'}}} 

\newcommand{\vXW}{\vec{\mathbf{X}}_{\W}} 
\newcommand{\vXWc}{\vec{\mathbf{X}}_{\Wc}} 
\newcommand{\vxW}{\vec{\mathbf{x}}_{\W}} 
\newcommand{\vxWc}{\vec{\mathbf{x}}_{\Wc}} 

\newcommand{\x}{\mathbf{x}} 
\newcommand{\X}{\mathbf{X}} 
\newcommand{\xW}{\mathbf{x}_{\W}} 
\newcommand{\XW}{\mathbf{X}_{\W}} 
\newcommand{\xWc}{\mathbf{x}_{\Wc}} 
\newcommand{\XWc}{\mathbf{X}_{\Wc}} 

\newcommand{\calX}{\mathcal{X}} 
\newcommand{\calXW}{\mathcal{X}_{\W}} 
\newcommand{\calXWc}{\mathcal{X}_{\Wc}} 

\newcommand{\K}{K} 
\newcommand{\vk}{\vec{\mathbf{k}}} 

\newcommand{\pinn}[1]{p^i_{\vX}\left( #1 \right)} 
\newcommand{\pact}[1]{p^a_{\vX}\left( #1 \right)} 
\newcommand{\phat}[1]{\hat{p}_{\vX}\left( #1 \right)} 

\newcommand{\pscal}[1]{p_{\X}\left( #1 \right)} 
\newcommand{\pscalW}[1]{p_{\XW}\left( #1 \right)} 

\newcommand{\pinnscal}[1]{p^i_{\X}\left( #1 \right)} 
\newcommand{\pinnscalW}[1]{p^i_{\X_{\W}}\left( #1 \right)} 

\newcommand{\phatscal}[1]{\hat{p}_{\X}\left( #1 \right)} 

\newcommand{\pinnm}[1]{p^i_{\vX_{\W}}\left( #1 \right)} 
\newcommand{\pactm}[1]{p^a_{\vX_{\W}}\left( #1 \right)} 
\newcommand{\phatm}[1]{\hat{p}_{\vX_{\W}}\left( #1 \right)} 

\newcommand{\pactscal}[1]{p^a_{\X}\left( #1 \right)} 
\newcommand{\pactscalW}[1]{p^a_{\X_{\W}}\left( #1 \right)} 
\newcommand{\pactscalWWc}[1]{p^a_{\X_{\W},\X_{\Wc}}\left( #1 \right)} 
\newcommand{\pactscalcond}[2]{p^a_{\X_{\Wc} | \X_{\W}}\left( #1 | #2\right)} 

\newcommand{\pstoch}[2]{p^{\textrm{stoch}}_{\vX|\vM}\left( #1 | #2\right)} 
\newcommand{\pstochH}[2]{p^{\textrm{stoch}}_{\vX|\vM}\left( #1 | #2\right)} 

\newcommand{\Bin}[1]{\mathcal{B}\left(#1\right)} 
\newcommand{\expect}[1]{\mathbb{E}_{\mathcal{B}}\left[ #1 \right]} 

\newcommand{\typ}{\mathcal{T}^{(n)}_{\eps}} 
\newcommand{\typevec}{Q} 
\newcommand{\typeclass}{\mathcal{Q}} 
\newcommand{\typeveccond}{Q_{\vxWc|\vxW}} 

\newcommand{\typcond}[1]{\mathcal{T}^{(n)}_{\eps}\left(#1\right)} 

\newcommand{\N}[2]{N\left(#1 | #2\right)} 

\newcommand{\eps}{\epsilon}
\newcommand{\epsden}{\epsilon_d} 
\newcommand{\epshid}{\epsilon_h} 
\newcommand{\epsrel}{\epsilon_r} 
\newcommand{\epssec}{\epsilon_s} 
\newcommand{\const}{\tilde{r}_d} 
\newcommand{\constone}{\tilde{r}_d^1} 
\newcommand{\epsone}{\epsilon_1}

\newcommand{\epsn}{\epsilon_n} 
\newcommand{\mB}{\hat{m}_B} 

\usepackage{framed}
\usepackage{color}

\begin{document}
%
\title{Reliable, Deniable, and Hidable Communication over Multipath Networks}
%
%
%

\author{Swanand~Kadhe, Sidharth~Jaggi, Mayank~Bakshi, and Alex~Sprintson 
\thanks{Swanand Kadhe and Alex Sprintson are with the Department
of Electrical and Computer Engineering, Texas A\&M University, College Station,
TX, 77840 USA; e-mails: \{kswanand1, spalex\}@tamu.edu.}
\thanks{Sidharth Jaggi is with the Department of Information Engineering, The Chinese University of Hong Kong, New Territories, HK; e-mail: jaggi@ie.cuhk.edu.hk}
\thanks{Mayank Bakshi is with the Institute of Network Coding at the Chinese University of Hong Kong, New Territories, HK; e-mail: mayank@inc.cuhk.edu.hk}
}

%
%

\markboth{IEEE Transactions on Information Theory}%
{To be Submitted}
%

\maketitle

\begin{abstract}
We consider the scenario wherein Alice wants to (potentially) communicate to the intended receiver Bob over a network consisting of multiple parallel links in the presence of a passive eavesdropper Willie, who observes an unknown subset of links. A primary goal of our communication protocol is to make the communication ``deniable'', {\it i.e.}, Willie should not be able to {\it reliably} estimate whether or not Alice is transmitting any {\it covert} information to Bob. 
Moreover, if Alice is indeed actively communicating, her covert messages should be information-theoretically ``hidable'' in the sense that Willie'€™s observations should not {\it leak any information} about Alice's (potential) message to Bob -- our notion of hidability is slightly stronger than the notion of information-theoretic strong secrecy well-studied in the literature, and may be of independent interest. 
It can be shown that deniability does not imply either hidability or (weak or strong) information-theoretic secrecy; nor does any form of information-theoretic secrecy imply deniability. We present matching inner and outer bounds on the capacity for deniable and hidable communication over {\it multipath networks}.
\end{abstract}


\IEEEpeerreviewmaketitle


\section{Introduction}
\label{sec:intro}

\IEEEPARstart{T}{he} urge to communicate, to speak and be heard, is a fundamental human need. However, embedded within our increasingly sophisticated communication networks, Big Brother is often watching. There are situations where even the fact that communication is happening (not just the content of that communication), can have real-world consequences. For instance, if you are a politically active citizen in an authoritarian society with broad censorship powers, the mere fact that you are communicating with the outside world can be construed by those authorities as sufficient justification for reprisals.

The goal of this paper is to investigate a class of communication models with a threefold objective. Firstly, all communication from the source (Alice) to the destination (Bob) should be reliable, {\it i.e.}, Bob should be able to identify the messages indented for him and decode them with high accuracy. Secondly, if the communication is overheard by a third party (Willie), it should be deniable from Willie. That is, Willie should not even be able to reliably decide whether or not Alice is indeed communicating with Bob. Finally, the communication should be hidable from Willie, {\it i.e.}, the eavesdropper Willie should not be able to learn anything about Alice's messages to Bob. Throughout the paper, we assume that Alice and Bob do {\it not} share any secret information that is not know to Willie.

Specifically, we consider the model wherein Willie is aware of Alice's ``innocent'' communication patterns (when she is not communicating covertly with Bob). However, due to resource limitations, Willie cannot wiretap on all of Alice's communication links, but only some of them. Willie, then, wishes to estimate whether Alice's communication is routine, or malevolent. Furthermore, Willie is also interested in inferring some information about Alice's (potential) covert communication with Bob. 

For example, perhaps during the course of a workday, highly placed government official Alice sends text messages, makes phone calls, writes letters, and posts on various websites. Intelligence analyst Willie, who is suspicious that Alice is perhaps a foreign spy, is keeping tabs on some of this activity, but, critically, is unable to see all of it. Willie's first goal is to differentiate between an ``innocent'' Alice (who communicates, in a somewhat predictable manner, with the outside world as a part of her daily life) and an ``active'' Alice (who is deliberately leaking secrets to spymaster Bob who gets to observe  Alice's transmissions on all channels). Willie's second goal is to infer some information about the secrets that Alice might be leaking to Bob.

Essentially, to be deniable, an active Alice has to look innocent in any subset of her communication links (since she does not know which of her communications might be wiretapped by suspicious authorities), {\it i.e.}, her communication on any subset of channels should be commensurate with what a normal person in Alice's position would do. Her covert communication with Bob, then, must be a function of all these channels. The challenge for Alice is to be able to embed meaningful information to Bob on channels which individually look innocent (deniability), and in such a way that Willie can infer absolutely nothing about her covert communication with Bob (hidability).

It is important to note that the conventional means of achieving secrecy, like cryptographic security, are not helpful in achieving the goal of deniability. On the contrary, if Willie finds that the data being communicated is encrypted, it can arouse his suspicion that Alice is {\it active}. 

Similarly, at first glance, it seems like a deniable scheme, which prevents Willie from estimating whether or not Alice is covertly communicating, will inherently be hidable (or secure). However, we demonstrate that the deniability and the hidability conditions are independent of each other. 

\textbf{Our contributions} can be summarized as follows. First, we formulate a mathematical model describing Alice's innocent and active communication patterns for a network with multiple parallel paths, referred to as the {\it multipath network}. Then, we formally define the notions of reliability, deniability, and hidability. For deniability, we use a hypothesis testing based metric that was used in \cite{BasGT:12}, \cite{CBJ:13}. Our condition of hidability is slightly stricter than the condition of strong information-theoretic secrecy (which essentially requires that the mutual information between the covert message and Willie's observation is bounded below a small constant, see \cite{MauW:00}), hence we use the term hidability rather than (information-theoretic) secrecy. It can be shown that hidability always guarantees the strong information-theoretic secrecy, but the converse is not necessarily true. 

Secondly, we characterize the capacity for reliable {\it and} deniable communication over a multipath network (referred to as {\it reliable-deniable} capacity). In our achievable strategy, we use random binning to generate the codebook and employ a (one-to-many) stochastic mapping to encode the covert messages. Finally, we characterize the capacity with the added requirement of hidability, and show that the random binning based stochastic encoding can also achieve the {\it reliable-deniable-hidable capacity}. 

The information-theoretic techniques that we use enable us to attain separability between the deniable encoding and the hidable encoding. This essentially means that the communication can be made either deniable (but not hidable) or hidable (but not deniable), or both deniable and hidable.

\section{Related Work}
\label{sec:related-work}
{\it Cryptography :} Even though cryptography allows communication at a rate higher than that possible by using other techniques, cryptographic techniques are not inherently deniable. Essentially, most cryptographic techniques make it computationally hard for an eavesdropper to distinguish the output (of the cryptographic system) from the output of a uniformly random noise sequence. Therefore, to achieve deniability, the outputs of most cryptographic schemes would still need to be ``shaped'' via techniques described in this paper.

{\it Information-theoretic secrecy :} At first sight, the proposed notions of deniability and hidability seem to have significant overlap with the notions of information-theoretic secrecy (see, {\it e.g.}, \cite{Mau:93}, \cite{MauW:00}, \cite{CsisK:78}). However, as we show in appendix~\ref{sec:hidability-vs-secrecy}, the proposed condition of hidability is a stricter condition than (strong) information-theoretic secrecy. Furthermore, we also show (in appendix~\ref{sec:deniability-vs-hidability}) that the deniability and the hidability are independent of each other (neither one implies the other).

{\it Network Anonymity :} Anonymizing protocols, such as Tor networks \cite{MBG:08}, enable users to route packets through crowds so that it is hard for eavesdroppers to estimate the source or the destination of the traffic. However, most anonymizing protocols (see {\it e.g.}, \cite{SPLB:08}) are not useful if Willie is eavesdropping at Alice's very point of connection to the network. For deniability, the active covert communication should look like an innocent behavior, and the very fact that Alice chooses to route packets through Tor might arouse Willie's suspicion.

{\it Steganography :} involves hiding messages into transmitted data such as images. The ideas presented in this paper come close to the information-theoretic steganography framework proposed in \cite{Cac:98}, \cite{WanM:08}, \cite{RyaR:09}. However, the main difference is that these steganographic protocols require (large) private keys to be shared between Alice and Bob, unlike our model, which requires no shared secret. But, these steganographic protocols allow the eavesdropper to observe the entire network, while  our model imposes restrictions on Willie's observation power.

{\it Other Deniable Protocols :} Our work falls in the line of \cite{BasGT:12} and \cite{CBJ:13}. In \cite{BasGT:12}, the authors show that deniability can be achieved (under AWGN channel model) if Alice can secretly share her codebook with Bob. Che {\it et al.} \cite{CBJ:13} show that any shared secrete is not necessary (under binary symmetric channel model), if the channel between Alice and Willie is noisier than that between Alice and Bob. Even though we use some information-theoretic techniques from \cite{CBJ:13} in this work, there are several key differences. First, in the model considered by \cite{CBJ:13} (and also \cite{BasGT:12}), in the innocent state, Alice opts to keep quiet (transmits the all zero codeword). On the other hand, in our work, we utilize Alice's innocent communication patterns to hide the covert messages. Secondly, in \cite{CBJ:13}, the authors exploit the fact that Willie's channel is noisier than Bob's to achieve deniability (thus, hiding the covert data in noise), we exploit Willie's limited observing power to be deniable (thus, hiding the covert data in different channels). Finally, these papers only focus on deniability and do not consider the hidability metric. 

{\it Hou and Kramer's work~\cite{HouK:13} :} comes closest to our work. We summarize the key similarities and the differences below.
\begin{itemize}
\item The notion of {\it effective secrecy} proposed by Hou and Kramer essentially combines together the metrics of deniability (referred to as {\it stealth} in \cite{HouK:13}) and hidability (\cite{HouK:13} considers strong information-theoretic secrecy). Therefore, the encoding strategies considered in \cite{HouK:13} always attain deniability and secrecy together. On the contrary, our stochastic encoding uses a ``low rate randomness'' for deniability, and a ``high-rate randomness'' for hidability. This flexibility allows us to get either deniability or hidability or both.\footnote{In principle, Hou and Kramer \cite{HouK:13} could have designed separate encoding schemes for deniability and secrecy, but their choice of metric, which merges the two conditions, leads to the single coding scheme achieving the both.} 
\item Hou and Kramer consider the generic wiretap channel model. Our model, in which Willie can observe an unknown subset of links, can be considered as a model of an arbitrarily varying channel (AVC)~\cite{CsisN:88}, where the capacity of the channel is a function of the subset that Willie is tapping.
\item While the capacity result of Hou and Kramer is more general from the perspective of the channel model, we specialize the capacity for deniable and hidable communication for the specific multipath network channel that we care about, and hence our capacity expression is explicitly computed, rather than corresponding to a convex optimization problem over an infinite alphabet.   
\end{itemize}

\section{Problem Formulation}
\label{sec:model}

\subsection{Notational Conventions}
\label{sec:notation}
Boldface upper-case symbols, {\it e.g.} $\mathbf{X}$, denote random variables, boldface lower-case symbols, {\it e.g.} $\mathbf{x}$, denote particular realizations of those random variables. Calligraphic symbols like $\W$ denote sets. The size of a set $\W$ is denoted as $\left|\W\right|$ and the complement is denoted as $\Wc$. Boldface symbols with arrow on their top, {\it e.g.}, $\vx$, denote vectors. We also use boldface symbols with an arrow on top to denote binary matrices. The distinction as to whether such symbol represents a vector or a binary matrix will be made clear. The reason behind using the same notation for vectors and binary matrices is that, in the latter part of the paper, we consider $m\times n$ binary matrices as length-$n$ vectors of symbols chosen from the finite field $GF(2^m)$. Unless otherwise specified, all vectors are of length $n$, where $n$ corresponds to the {\it block-length} (number of network uses). The probabilities of the events are denoted by symbol $\Pr$ with a subscript denoting the random variable(s) over which the probabilities are calculated. 

We use standard notation for information-theoretic quantities: $\Hp{.}$ denotes the entropy function, $\Hcond{.}{.}$ denotes the conditional entropy function, $\I{.}{.}$ denotes the mutual information, and $\D{.}{.}$ denotes the Kullback-Leibler divergence between two distributions. Lastly, we use the notation $\Vdist{p_1(.)}{p_2(.)}$ to denote the total variation distance between any two distributions $p_1(.)$ and $p_2(.)$, defined over an alphabet $\calX$, defined as follows: 
\begin{equation}
\label{eq:var-dist}
\Vdist{p_1(.)}{p_2(.)} = \frac{1}{2} \sum_{\x \in \calX} \left\vert {p_1(\x) - p_2(\x)}  \right\vert.
\end{equation}  

\subsection{Problem Statement}
\label{sec:problem}
Suppose Alice wants to (potentially) communicate with Bob over a {\it multipath} network consisting of $\C$ parallel links. Each link is assumed to have the capacity of one bit per use\footnote{Note that, if the links have unequal capacities, it is possible to split each link into multiple links with the same capacity.}
and is assumed to be noiseless\footnote{It is worth noting that i.i.d. noise on links does not fundamentally weaken our results. In this case, one can use link-by-link error correction. However, in this paper, we do not explore along this direction.}. Alice is allowed to transmit a length-$n$ binary sequence $\vx_i\in\{0,1\}^{1\times n}$ on the $i$-th link (here, $i$ is an index in $\{1,\ldots,\C\}$) over $n$ network uses (here, $n$ is the {\it block-length}). Bob receives the set of $\C$ sequences and organizes them as the $\C \times n$ binary received {\it codeword matrix} as $\vx = \left[\vx_1^T\: \vx_2^T\: \cdots \: \vx_{\C}^T \right]^T$.    

\begin{figure}[!t]
 \centering
  \def\svgwidth{600pt}
  \small{
  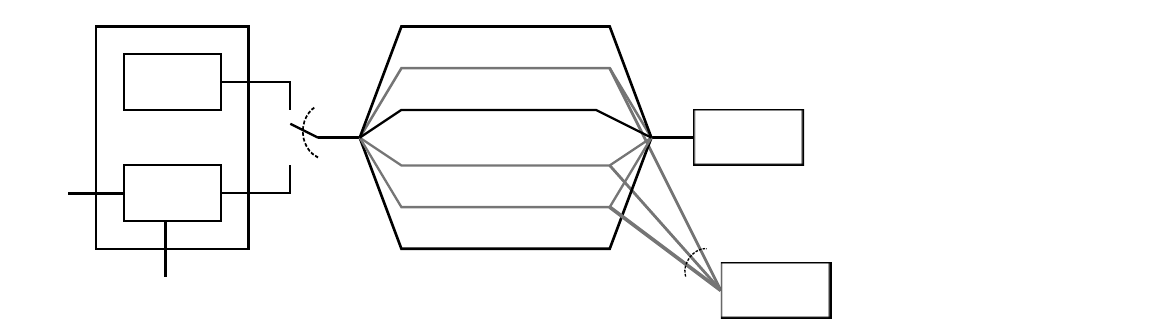
  }
  \caption{System model for reliable, deniable, and hidable communication over a multipath network. For reliability, Bob's error probability should be bounded, for deniability Willie should not be able to distinguish between active and innocent Alice, and for hidability Willie should not be able to infer anything about Alice's covert messages.}
\end{figure}


Alice may or may not wish to communicate covertly with Bob, and accordingly she is said to be in an {\it active} or in an {\it innocent} state. If Alice does not have any covert message to transmit to Bob, she is said to be in an innocent state. In the innocent state, during the $t$-th time instant $(1\leq t\leq n)$, Alice randomly generates a length-$\C$ binary {\it codeword} $\x(t) = \left[\x_1(t) \: \x_2(t) \: \cdots \: \x_\C(t)\right]^T$, denoting the transmissions on each of the links, according to the distribution $p^i_{\X(t)}(\x(t))$ (over an alphabet of size $2^{\C}$) called as the {\it scalar innocent distribution}. We assume that the codewords $\x(t)$ are independent and identically distributed (i.i.d.) over the time instants $t$, $1\leq t\leq n$.\footnote{One can question this assumption that the binary vectors transmitted over the multipath network are i.i.d. over the time instants $t$, $1\leq t\leq n$. However, we should point out that getting the results under this (i.i.d. assumption) model is still challenging; we aim to extend this model to more realistic source models in the future work.} Therefore, the $\C \times n$ binary codeword matrix $\vx$ that is transmitted by Alice is distributed according to the distribution $\pinn{\vx} = \pinnscal{\x(1)}\pinnscal{\x(2)}\cdots \pinnscal{\x(n)}$, wherein $\vx = \left[\x(1) \: \x(2) \cdots \x(n)\right]$. This distribution $\pinn{.}$, defined over the alphabet of all binary $\C \times n$ binary matrices, is called as the {\it innocent distribution}.


When Alice is in the active state, she wants to transmit a {\it covert message} $\M \in \left\{1, \ldots, 2^{nR}\right\}$ to Bob. We assume that the message symbols are distributed uniformly. Let $\vM$ be the random variable corresponding to the length-$nR$ binary vector representing the covert message -- here, $R$ denotes the {\it rate} of Alice's covert message. Alice encodes her covert messages using an {\it encoder} $\enc : \{0,1\}^{nR} \times \{0,1\}^{nr} \rightarrow \{0,1\}^{\C \times n}$, where $r$ denotes the {\it rate of private randomness} used by her encoder. 
The set $\{\vx_1,\ldots,\vx_{|\code|}\}$ of all the possible output codeword matrices of the encoder $\enc$ forms the {\it codebook} $\code$ for Alice. Notice that Alice's encoder induces a distribution on the transmitted binary codeword matrices. This distribution $\phat{.}$, defined over the alphabet of all binary $\C \times n$ matrices, that is induced by Alice's encoding process in the active state is called as the {\it induced distribution}.

We use a binary random variable $\T$ to denote Alice's transmission status, with $\T = 0$ means Alice is innocent and $\T = 1$ means she is active. We assume that the prior statistics on $\T$ can be known to Willie, but need not be known to Bob. Further, we assume that only Alice knows the value of $\T$ \emph{a priori}. 

Communication takes place in the presence of a passive eavesdropper Willie, who can observe some subset of links. Let $\W$ denote the set of links that are eavesdropped by Willie, and $\bigW$ denote the class of all possible subsets of links which Willie can observe. For example, $\bigW$ might comprise of all subsets of at most $\C - 1$ links. Let $\vxW\in\{0,1\}^{\left|\W\right|\times n}$ be the codeword (sub-)matrix that is observed by Willie. The marginal distribution on the codewords that Willie observes when Alice is innocent, denoted as $\pinnm{.}$, defined over the alphabet of all the binary $\left|\W\right| \times n$ matrices, is called as the {\it marginal innocent distribution}. Similarly, the marginal distribution on the codewords observed by Willie when Alice is active, denoted as $\phatm{.}$, defined over the alphabet of all binary $\left|\W\right| \times n$ matrices, is called as the {\it marginal induced distribution}. 

Throughout, we assume that there is {\it no} shared secrete between Alice and Bob that is not known to Willie. Further, we assume that Willie knows the encoding-decoding scheme used by Alice and  Bob.

\begin{figure}[!t]
 \centering
  \subfigure[System Diagram]{
  \label{fig:ex-1-a}
  \def\svgwidth{450pt}
  \small{
  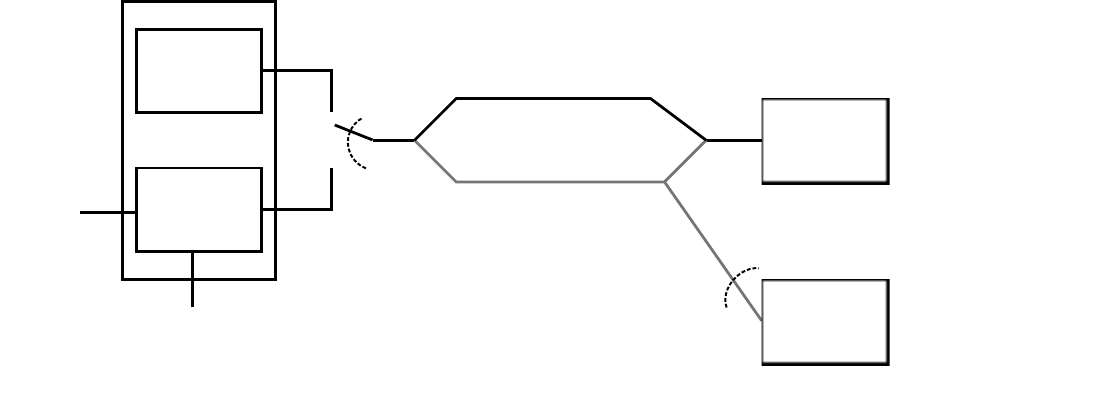
  }
  } \\
  \subfigure[Scalar Innocent Distribution]{
  \label{fig:ex-1-b}
  \begin{tabular}{|c|c|}
  \hline
  $\X(t) = \left[\X_1(t)\: \X_2(t)\right]^T$ & $\pinnscal{.}$ \\
  \hline
  $0\: 0$ & 0 \\
  $0\: 1$ & $\frac{1}{2}$\\
  $1\: 0$ & $\frac{1}{4}$\\
  $1\: 1$ & $\frac{1}{4}$\\
  \hline
  \end{tabular}} \qquad \qquad
  \subfigure[Scalar Induced Distribution]{
  \label{fig:ex-1-c}
  \begin{tabular}{|c|c|}
  \hline
  $\X(t) = \left[\X_1(t)\: \X_2(t)\right]^T$ & $\phatscal{.}$ \\
  \hline
  $0\: 0$ & $\frac{1}{4}$\\
  $0\: 1$ & $\frac{1}{4}$\\
  $1\: 0$ & $\frac{1}{4}$\\
  $1\: 1$ & $\frac{1}{4}$\\
  \hline
  \end{tabular}}
  \caption{Toy Example of a multipath network with $\C = 2$ links. Willie can observe any one of the links.} 	
  \label{fig:ex-1}
\end{figure}

{\it Toy Example:} Consider a multipath network consisting of $\C = 2$ links as shown in Fig.~\ref{fig:ex-1}. Suppose that Willie can eavesdrop on any one of the links. Thus, $\bigW = \left\{\{1\},\{2\}\right\}$. When innocent, at each time instant $t$, $1\leq t\leq n$, Alice choses length-$2$ binary codewords according to the scalar innocent distribution given in Fig.~\ref{fig:ex-1-b}. Note that the innocent distribution on binary $2\times n$ codeword matrices will be the product distribution given by $\pinn{.} = \{p^i_{\X}\}^n$. The marginal innocent distribution on the top link will be the Bernoulli process with parameter $\frac{1}{2}$, while the marginal innocent distribution on the bottom link will be the Bernoulli process with parameter $\frac{3}{4}$.

Suppose Alice wants to transmit a binary, length-$n$ covert message $\vM \in \{0,1\}^n$ to Bob. Notice that this results in Alice's transmission rate to be $R = 1$. Suppose that she uses Shannon's one-time padding scheme as her stochastic encoder, described as follows. For each time instant $t$, Alice generates a uniform random bit $\K(t)$. On the top link, she transmits the random bit $\K(t)$, whereas, on the bottom link, she transmits exclusive OR of the covert message bit and the random bit $\M(t) \oplus \K(t)$. With this encoding, the induced distribution $\phat{.}$ will be the uniform distribution over all the binary $2\times n$ matrices. The marginal induced distribution on each link will also be the uniform distribution over the length-$n$ binary vectors.    

We aim to design a communication scheme that satisfies the following three requirements:

\emph{1) Reliability} : Bob should be able to estimate whether Alice is active or innocent and decode the messages correctly when Alice is active, with high probability over Alice's encoding scheme. This is, in fact, a basic requirement for any reasonable communication scheme, but we make it explicit here, and formally define this below.
\begin{definition}[Reliability]
We say that the scheme is $(1-\epsrel)$-reliable if, for any arbitrarily small $\epsrel > 0$,
\begin{equation}
\label{eq:reliability}
\Pr{}_{\vX}\left({\TB = 1 | \T = 0}\right) + \Pr{}_{\vX}\left({\TB = 0 | \T = 1}\right) + \Pr{}_{\vM,\enc}\left({\vMB \neq \vM} | \T = 1\right) \leq \epsrel,
\end{equation}
where $\TB$ and $\vMB$ denote Bob's estimates of Alice's transmission status $\T$ and her covert message $\vM$, respectively.
\end{definition}
The first term in the summation in~\eqref{eq:reliability} gives the probability, over the randomness of the codeword matrices, that Bob estimates Alice to be active conditioned on the fact that she, actually, is innocent; the second term is the probability, over the randomness of the codeword matrices, that Bob finds Alice to be innocent given that she is indeed active; and the last term is the conditional probability, over the randomness of the covert messages and the private randomness of Alice's encoder, given that Alice is active, that Bob wrongly decodes the covert message. Notice that each of these three events is an error event, and for the scheme to be reliable, the probability of each of the error events should be very small, as defined above. 

\emph{2) Deniability} : Willie should not be able to ``reliably'' estimate Alice's transmission status $\T$, as formally defined below.

\begin{definition}[Deniability] 
We say that the scheme is $(1-\epsden)$-deniable if, for any arbitrarily small $\epsden > 0$,
\begin{equation}
\label{eq:deniability}
\Vdist{\pinnm{.}}{\phatm{.}} < \epsden, \quad \forall \: \W \in \bigW,
\end{equation}
where $\Vdist{.}{.}$ denotes the total variation distance defined in~\eqref{eq:var-dist}.
\end{definition}
To get some intuition behind the definition of deniability, notice that Willie essentially performs binary hypothesis testing on the parameter $\T$. One can show, by standard hypothesis testing arguments \cite{LehR:05}, that if $\Vdist{\pinnm{.}}{\phatm{.}}$ is upper bounded by some small $\epsden$, then Willie's best estimator based on his observations is at most $\epsden$ better than even a na\"ive estimator independent of his network observations.

\emph{3) Hidability} : When Alice is active, Willie's observations on the links in $\W$ should, with high probability, not ``leak any information'' about the message Alice is transmitting, as defined in Definition 3 below.

\begin{definition}[Hidability] 
We say that the scheme is $(1- \epshid)$-hidable if, for any arbitrarily small $\epshid$, $0< \epshid <1$, we have  
\begin{equation}
\label{eq:hidability}
1 - \epshid \leq \frac{\Pr{}_{\vM,\enc}\left(\vM = \vm | \vxW, \T = 1\right)}{\Pr{}_{\vM}\left(\vM = \vm | \T = 1 \right)} \leq 1 + \epshid, \quad \forall \: \vm, \: \forall \: \W \in \bigW,
\end{equation}
for any binary $\left|\W\right| \times n$ matrix $\vxW$ that Willie observes.
\end{definition}
Note that our definition of hidability is somewhat stronger than the usual definition of information-theoretic secrecy, since, as we show in Appendix~\ref{sec:hidability-vs-secrecy}, $(1-\oh{\frac{1}{n}})$-hidability implies information-theoretic secrecy, but the converse is not necessarily true. Furthermore, at first sight, it seems like the deniability is a stronger condition than the hidability because deniability requires that the transmission status $\T$ should be indistinguishable to Willie irrespective of whether he can decode the codewords or not, but actually neither implies the other (see Appendix~\ref{sec:deniability-vs-hidability} for examples of hidable schemes that are not deniable, and deniable schemes that are not hidable.)

\begin{table}[!t]
\begin{center}
\begin{tabular}{|l| |l|}
\hline
$\T$ & Alice's transmission status \\
$\TB$ & Bob's estimate of Alice's transmission status \\
$\TW$ & Willie's estimate of Alice's transmission status \\
$\C$ & Number of parallel links in the network \\
$\bigW$ & Class of all possible subsets of links which Willie can observe \\
$\W$ & Subset of links observed by Willie \\
$M$ & Number of messages Alice wishes to transmit when active\\
$\vm$ & Particular message transmitted by Alice (n-bit long) \\
$\vM$ & Random variable (as a binary vector) corresponding to message \\
$\vMB$ & Bob's estimate of the transmitted message \\
$\vMW$ & Willie's estimate of the transmitted message \\
$\vx$ & Codeword transmitted by Alice \\
$\vX$ & Random variable corresponding to codeword \\
$\calX$ & Alphabet for codeword symbol \\
$\pinnscal{\x}$ & Innocent distribution on codeword symbols \\
$\pinn{\vx}$ & Innocent distribution on n-symbol length codeword \\
$\phat{\vx}$ & Distribution induced by Alice's encoding in active state \\
$\vxW$ & Codeword observed by Willie \\
$\vXW$ & Random variable corresponding to the codeword observed by Willie \\
$\calXW$ & Alphabet for codeword symbols observed by Willie \\
$\pinnm{\vxW}$ & Marginal innocent distribution observed by Willie \\
$\phatm{\vxW}$ & Marginal active distribution observed by Willie \\
$\vxWc$ & Codeword that Willie cannot observe \\
$\calXWc$ & Alphabet for codeword symbols that Willie cannot observe \\
$\typ$ & Strongly typical set of codewords \\
$\typevec$ & Type of a codeword and also type class corresponding to that type \\
$\typcond{\vXWc | \vxW}$ & Conditionally strongly typical set \\ 
\hline
\end{tabular}
\caption{Notation used throughout the paper}
\end{center}
\end{table}

\section{Deniable Encoders}
\label{sec:deniable-encoders}
In this section, we focus on designing the schemes that achieve only deniability. In further sections, we extend these schemes to be hidable as well as deniable.

\subsection{Main Result for Deniable Communication}
\label{sec:deniable-main-result}
We assume that the scalar induced distribution $\pinnscal{.}$, defined over the alphabet $\calX = \left\{0,1,\ldots,2^{\C}-1\right\}$, is given. This results on the innocent distribution $\pinn{.}$ as the product distribution $\pinn{.} = \prod_{t=1}^n\pinnscal{.}$ (under the assumption that the codewords are i.i.d. over time instants). We want to know what is the maximum rate at which Alice can transmit the covert messages and how she can perform encoding such that her active status is simultaneously reliable to Bob and deniable from Willie. 

Before characterizing the {\it reliable-deniable} capacity, we need to review several information-theoretic concepts related to {\it typicality} (see \cite{CsisK-book:2011} for details). Hereafter, we consider the $\C$ parallel links as one hyperlink with input (output) alphabet as $\calX = \left\{0,1,\ldots,2^{\C}-1\right\}$. Thus, all the codewords will be length-$n$ vectors with each symbol belonging to alphabet $\calX$. 

Let us begin with the notion of type. The {\it type} $\typevec$ of a sequence $\vx\in\calX^n$ is the empirical distribution on $\calX$ defined as \cite{CsisK-book:2011}:
\begin{equation}
\label{eq:type-vector}
\typevec_{\vx} = \frac{1}{n}N\left(a | \vx\right) \quad \forall a\in\calX,
\end{equation}
where $\N{a}{\vx}$ denotes the number of occurrences of a symbol $a\in\calX$ in the sequence $\vx$. The set of all sequences of type $\typevec$ is called as {\it type class} $\typevec$. For brevity, we use the same notation to denote both the type and the type class corresponding to that type; the distinction will be clear from the context. 

For some small $\eps > 0$, the $\epsilon${\it -strongly typical set} (w.r.t. a distribution $\pscal{.}$) is defined as follows \cite{CsisK-book:2011}:
\begin{equation}
\label{eq:typ-set-C-ary}
\typ(\pscal{.}) = 
\left\{\vx : 
\begin{array}{ll} 
\left| \frac{1}{n} \N{a}{\vx} - \pscal{a} \right| \leq \frac{\eps}{\left|\calX\right|} \quad \forall a \in \calX, & \textrm{if} \quad \pscal{a} > 0\\
\N{a}{\vx}  = 0, & \textrm{if}  \quad \pscal{a} = 0
\end{array}
\right\}.
\end{equation}
Note that the typical set can be viewed as a collection of the type classes, which satisfy the aforementioned constraint. It is worth noting that the probability of each sequence in a particular type class is the same.

In the following we present the matching inner and outer bounds on the rate of covert transmission for reliable-deniable communication.
\begin{theorem}
\label{thm:deniability-multilink}
The capacity of the {\it reliable-deniable} communication over a multipath network is 
\begin{equation}
\label{eq:deniable-rate}
\Capd = \sup_{\substack{\pscal{.} : \forall\W\in\bigW \\ \pscalW{.} = \pinnscalW{.}}} \Hp{\pscal{.}}.
\end{equation}
In other words, for any sufficiently small $\delta$, any scalar innocent distribution $\pinnscal{.}$, any covert transmission rate $\Rd < \Capd$, there exists an encoder $\enc : \{0,1\}^{n\Rd} \times \{0,1\}^{nr} \to \{0,1\}^{\C \times n}$ that is simultaneously $(1-\epsrel)$-reliable and $(1-\epsden)$-deniable for any $0 \leq \epsrel, \epsden < \delta$ with high probability for sufficiently large block-length $n$. Conversely, any encoder with rate $\Rd \geq \Capd$ cannot be deniable.
\end{theorem} 

\subsection{Converse}
\label{sec:deniability-Converse}
We use standard information-theoretic arguments as follows. Given that Alice is active, we have
\begin{IEEEeqnarray}{rCl}
n\Rd & \leq & \Hp{\M}, \nonumber\\
& = & \I{\M}{\vX} + \Hcond{\M}{\vX}, \nonumber\\
& \stackrel{(a)} = & \I{\M}{\vX} + n\epsn, \nonumber\\
& \stackrel{(b)}{\leq} & \Hp{\vX} + n\epsn, \nonumber\\
& \stackrel{(c)}{\leq} & \sum_{j=1}^n \Hp{\vX(j)} + n\epsn, \nonumber\\
\label{eq:maximization-deniability}
& \stackrel{(d)}{\leq} & n \sup_{\substack{\pscal{.} : \forall\W\in\bigW \\ \pscalW{.} = \pinnscalW{.}}} \Hp{\pscal{.}} + n\epsn,
\end{IEEEeqnarray}
where $\epsn\to 0$ as $n\to\infty$, and (a) follows from Fano's inequality, (b) is due to non-negativity of entropy, (c) is due to the independence bound on entropy, and (d) due to the deniability condition.

\subsection{Achievability}
\label{sec:deniability-achievability}
Random Binning Based Stochastic Encoding:

\textbf{Codebook:}  Let $\pactscal{.}$ denote the distribution, defined over the alphabet $\calX := \{0,1,\cdots,2^{\C}-1\}$, that is used by Alice to generate her codeword symbols (referred to as the {\it scalar active distribution}). Alice computes $\pactscal{.}$ by solving the following convex optimization problem.
\begin{equation}
\label{eq:convex-opti}
\pactscal{.} = \argmax_{\substack{\pscal{.} : \forall\W\in\bigW \\ \pscalW{.} = \pinnscalW{.}}} \Hp{\pscal{.}}.
\end{equation}
Note that the codeword sequences will be distributed according to the {\it active distribution} $\pact{.} = \prod_{t=1}^n\pactscal{.}$, defined over the alphabet $\calX^n$. 

\begin{figure}[!t]
 \centering
  \def\svgwidth{0.6\columnwidth}
  \small{
  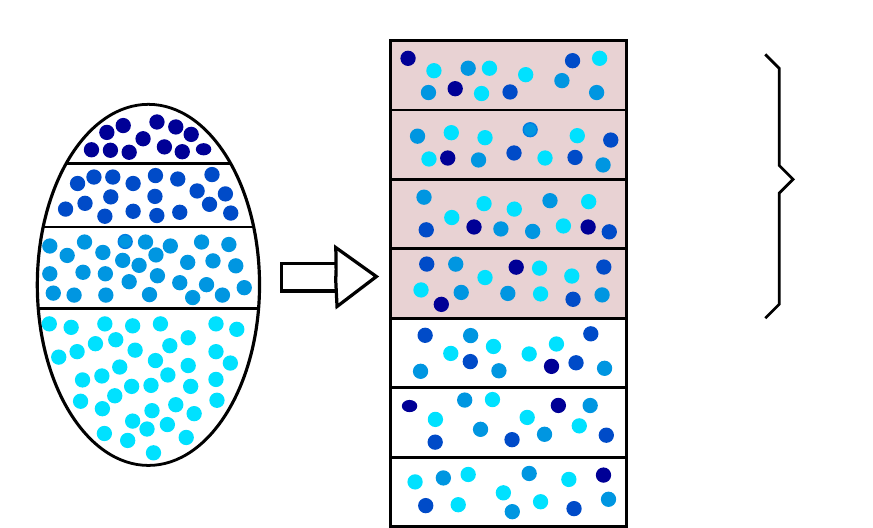
  }
  \caption{Random binning based codebook generation. The strongly typical set is shown as a collection of ``type classes'' $\typevec_j$, where all the sequences in a particular type class have the same empirical distribution (type).}
  \label{fig:binning}
\end{figure}

For some small $\eps > 0$, consider the $\eps$-\emph{strongly typical} set
\begin{equation}
\label{eq:typ-set-C-ary-1}
\typ\left(\pactscal{.}\right) = 
\left\{\vx : 
\begin{array}{ll} 
\left| \frac{1}{n} \N{a}{\vx} - \pactscal{a} \right| \leq \frac{\eps}{\left|\calX\right|} \quad \forall a \in \calX, & \textrm{if} \quad \pactscal{a} > 0\\
\N{a}{\vx}  = 0, & \textrm{if}  \quad \pactscal{a} = 0
\end{array}
\right\}.
\end{equation}
Randomly bin the sequences in the strongly typical set $\typ\left(\pactscal{.}\right)$ into $2^{n\R}$ bins, where $\R = \Hp{p^a_{\X}} - \Oh{\eps\log_2\frac{1}{\eps}+\frac{\log_2 n}{n}}$. Let $\Bin{j}$ denote the set of sequences that belong to bin $j$, $1\leq j \leq 2^{n\R}$. Generate the codebook $\code$ as the set of sequences that are fallen in the first $\frac{2^{n\R}}{n}$ bins, {\it i.e.}, 
\begin{equation}
\label{eq:codebook}
\code = \bigcup_{j=1}^{ 2^{n\Rd}} \Bin{j}, 
\end{equation}
where $\Rd = \R - \frac{\log_2 n}{n}$. Fig.~\ref{fig:binning} depicts the codebook generation process. Note that the covert transmission rate is $\Rd = \Hp{p^a_{\X}} - \Oh{\eps\log_2\frac{1}{\eps}+\frac{\log_2 n}{n}} -  \frac{\log_2 n}{n}$. 

\textbf{Encoding:} Associate each length-$n\R$, binary message vector $\vm$ with the $\m$-th bin $\Bin{\m}$, where $\m$ denotes the index corresponding to the binary message vector $\vm$. For simplicity, we denote the $m$-th bin interchangeably as $\Bin{\m}$ or $\Bin{\vm}$. 

To transmit a message $\vm$, randomly choose a codeword from $\Bin{\vm}$ according to the following conditional distribution.
\begin{equation}
\label{eq:stochastic-enc}
\pstoch{\vx}{\m} = 
\begin{cases}
\frac{\pact{\vx}}{\sum_{\vx:\vx\in\Bin{\m}}\pact{\vx}} & \textrm{if} \: \vx\in\Bin{\m},\\
0 & \textrm{Otherwise}.
\end{cases}
\end{equation}
Note that distribution $\pstoch{\cdot}{\cdot}$ defines a one-to-many stochastic encoder $\enc : \M \to \code$. Further, notice that the proposed way of generating the codebook and the stochastic mapping induces a particular distribution $\phat{\vx}$ on all $\vx\in\code$, which depends on $\pstoch{\vx}{\m}$.

\textbf{Decoding:} Let $\vx$ be the received codeword. If there exists a message $\tilde{\m}$ such that $\vx\in\Bin{\tilde{\m}}$, then declare that Alice is active and the decoded message $\mB = \tilde{\m}$; else declare that Alice is innocent.

\textbf{Deniability:} To prove deniability, we need to analyze the marginal distribution $\phatm{.}$ that Willie observes when Alice is active. To aid that, we first prove some key properties of the proposed encoding. The proofs can be found in Appendix~\ref{sec:useful-lemmas}.

First we show that the proposed encoding ensures that, for any type class that falls in the strongly typical set, the number of codewords of that particular type class per message bin is concentrated around its mean value, if the total number of bins considered in the encoding $(2^{n\R})$ is less than a particular threshold.
\begin{lemma}
\label{lem:concentration-typeclass}
In the random binning based stochastic encoding encoding, for any type class $\typevec\in\typ\left(\pactscal{.}\right)$, for $1\leq\m\leq 2^{n\R}$, we have
\begin{equation}
\label{eq:concentration-typeclass}
\Big| {\left\vert \typevec \cap \Bin{\m}\right\vert - \expect{\left\vert \typevec \cap \Bin{\m}\right\vert}} \Big| \leq \epsilon \: \expect{\left\vert \typevec \cap \Bin{\m}\right\vert}
\end{equation}
with high probability over binning process, provided provided 
$\R \leq \Hp{\pactscal{.}} - \const$, where $\const \geq \eps\log_2\frac{2^{\C}}{\eps} + \frac{2^{\C}\log_2(n+1)}{n}$.
\end{lemma}

Second, let us define the probability of a bin as the sum of probabilities (under active distribution) of all the codewords that have fallen in that particular bin. In particular, for a bin $\Bin{\m}$, $1\leq\m\leq 2^{n\R}$, $\Pr{}_{\mathcal{B}}\left(\Bin{\m}\right) := \sum_{\vx:\vx\in\Bin{\m}}\pact{\vx}$, where the probability is computed over the randomness of the binning process. The following lemma shows that the probability of each bin is roughly uniform with high probability.
\begin{lemma}
\label{lem:bin-prob-bound}
In the random binning based stochastic encoding, with high probability, the probability of a bin $\Pr{}_{\mathcal{B}}\left(\Bin{\m}\right) := \sum_{\vx:\vx\in\Bin{\m}}\pact{\vx}$, for $1\leq \m \leq 2^{n\R}$, is bounded as
\begin{equation}
\label{eq:bin-prob-bound}
{(1-\epsone)}\frac{1}{2^{n\R}} \leq \Pr{}_{\mathcal{B}}\left(\Bin{\m}\right) \leq {(1+\epsone)}\frac{1}{2^{n\R}}
\end{equation}
where $\epsone = \eps + (1-\eps)2^{\C+1}\exp{\left(-2\eps^2 n\right)}$, provided
\begin{equation}
\label{eq:super-rate-deniability}
\R \leq \Hp{\pactscal{.}} - \const, \qquad \textrm{where} \quad \const \geq \eps\log_2\frac{2^{\C}}{\eps} + \frac{2^{\C}\log_2(n+1)}{n}.
\end{equation}
\end{lemma}

Third, we show that, when Alice is active, the proposed encoding imposes certain typicality conditions on the sequences that are observed by (and also on the ones that are not observed by) Wille. 
\begin{lemma}
\label{lem:Willie-typicality}
Under random binning based stochastic encoding, the sequences $\vxW$ observed by Willie are $\eps$-strongly typical w.r.t. the marginal distribution $\pactscalW{\xW}$. Further, this in turn implies that the sequences $\vxWc$ that Willie cannot observe are $\frac{\eps\left(1+\left|\calXWc\right|\right)}{\left|\calXW\right|}$-conditionally strongly typical given $\vxW$ w.r.t. the conditional distribution $\pactscalcond{\xWc}{\xW}$.
\end{lemma}

Now, to prove deniability, let us consider the marginal induced distribution as follows.
\begin{IEEEeqnarray}{rCl}
\phatm{\vxW} & = & \sum_{\vxdash_{\Wc}}\phat{\vxdash_{\Wc},\vxW} = \sum_{\vxdash : \vxdash_{\W} = \vxW} \phat{\vxdash}\\
& \stackrel{(a)}{=} & \sum_{\m = 1}^{2^{n\Rd}} \sum_{\vxdash : \vxdash_{\W} = \vxW} \Pr{}_{\vM,\enc}(\vxdash,\m | \T=1) \\
& = & \sum_{\m = 1}^{2^{n\Rd}} \sum_{\vxdash : \vxdash_{\W} = \vxW} \Pr{}_{\vM,\enc}(\vxdash |\m, \T=1) \Pr{}_{\vM}(\m | \T=1)\\
\label{eq:marginal-induced}
& \stackrel{(b)}{=} & \sum_{\m = 1}^{2^{n\Rd}} \sum_{\substack{\vxdash : \vxdash_{\W} = \vxW \\ \vxdash\in\Bin{\m}}} \frac{\pact{\vxdash}}{\Pr_{\mathcal{B}}\left(\Bin{\m}\right)} \left(\frac{1}{2^{n\Rd}}\right),
\end{IEEEeqnarray}
where (a) follows from the law of total probability, and (b) follows from the definition of the stochastic encoding ({\it cf.}~\eqref{eq:stochastic-enc}) and that the covert messages are uniform over $2^{n\Rd}$ choices.

Since $\R$ satisfies the condition mentioned in~\eqref{eq:super-rate-deniability}, using~\eqref{eq:bin-prob-bound} in Lemma~\ref{lem:bin-prob-bound}, we have, with high probability, that
\begin{equation}
\label{eq:bounds}
\frac{2^{n(\R-\Rd)}}{1+\epsone}\sum_{1\leq\m\leq 2^{n\Rd}} \sum_{\substack{\vxdash : \vxdash_{\W} = \vxW \\ \vxdash\in\Bin{\m}}} \pact{\vxdash} \leq \phatm{\vxW}  \leq
\frac{2^{n(\R-\Rd)}}{1-\epsone}\sum_{1\leq\m\leq 2^{n\Rd}} \sum_{\substack{\vxdash : \vxdash_{\W} = \vxW \\ \vxdash\in\Bin{\m}}} \pact{\vxdash},
\end{equation}
where $\epsone = \eps + 2^{\C+1}\exp{\left(-2\eps^2 n\right)}\left(1-\eps\right)$.
Note that
$$ \sum_{1\leq\m\leq 2^{n\Rd}} \sum_{\substack{\vxdash : \vxdash_{\W} = \vxW \\ \vxdash\in\Bin{\m}}} \pact{\vxdash} \stackrel{(c)}{=} \sum_{\substack{\vxdash : \vxdash_{\W} = \vxW \\ \vxdash\in\bigcup_{1\leq\m\leq 2^{n\Rd}}\Bin{\vm}}}\pact{\vxdash} \stackrel{(d)}{=} \sum_{\substack{\vxdash : \vxdash_{\W} = \vxW \\ \vxdash\in\code}}\pact{\vxdash},$$  
where (c) follows because the sets $\Bin{\m}$ are disjoint for all $\m$, and (d) is due to the definition of the codebook ({\it cf.}~\eqref{eq:codebook}).

Using lemma~\ref{lem:Willie-typicality}, we can see that the set $\left\{\vxdash : \vxdash\in\typ, \vxdash_{\W}=\vxW\right\}$ is isomorphic to the conditionally strongly typical set $\typcond{\pactscalcond{.}{.}}$. Let $\codeW = \left\{\vxdash_{\Wc} : [\vxW; \vxdash_{\Wc}] \in \code\right\}$. Then, we can write
\begin{IEEEeqnarray}{rCl}
\sum_{\substack{\vxdash : \vxdash_{\W} = \vxW \\ \vxdash\in\code}}\pact{\vxdash} & = &\sum_{\vxdash : \vxdash_{\Wc} \in\codeW}\pact{\vxdash} \\
& = & \sum_{\typevec\in\typ(\vXWc | \vxW)}\sum_{\vxdash : \vxdash_{\Wc}\in\typevec\cap\codeW}\pact{\vxdash} \\
& \stackrel{(e)}{=} & \sum_{\typevec\in\typ(\vXWc | \vxW)}\left|\typevec\cap\codeW\right|\pact{\vxdash : \vxdash_{\Wc} \in {\typevec}} \\
& \stackrel{(f)}{\in} & \left[(1-\eps),(1+\eps)\right]
\sum_{\typevec\in\typ(\vXWc | \vxW)}\frac{\left|\typevec\right|}{n} \pact{\vxdash : \vxdash_{\Wc} \in {\typevec}} \qquad \textrm{w.h.p.} \\
& \stackrel{(g)}{\in} & \left[(1-\eps),(1+\eps)\right]  \frac{1}{n} \sum_{\substack{\vxdash : \vxdash_{\W} = \vxW \\ \vxdash\in\typ(\vX)}}\pinn{\vxdash} \quad \textrm{w.h.p.}
\end{IEEEeqnarray}
where (e) follows because for given $\vxW$, $\pact{\vxdash : \vxdash_{\Wc} \in {\typevec}}$ is the same for all $\left\{\vxdash : \vxdash_{\Wc} \in {\typevec}\right\}$, (f) is due to the fact that $\codeW$ can be considered as a {\it super-bin} composed of $2^{n\Rd}$ bins and then using the similar arguments as in Lemma~\ref{lem:concentration-typeclass} for concentration of the codewords of conditional type class\footnote{Note that the number of codewords of a conditional type class falling in the super-bin will be concentrated around its mean only if $\Hcond{\XWc}{\XW} > 0$. Also, see Remark~\ref{rem:corner-point-distributions}.}, and (g) follows due to the isomorphism between the sets $\left\{\vxdash : \vxdash\in\typ, \vxdash_{\W}=\vxW\right\}$ and $\typcond{\vXWc | \vxW}$.

Substituting back into~\eqref{eq:bounds} and by using $\Rd = \R - \log_2 n$, we have, with high probability, that
\begin{equation}
\label{eq:bounds-1}
\frac{(1-\eps)}{(1+\epsone)} \sum_{\substack{\vxdash : \vxdash_{\W} = \vxW \\ \vxdash\in\typ(\vX)}} \pact{\vxdash} \leq \phatm{\vxW}  \leq
\frac{(1+\eps)}{(1-\epsone)}\sum_{\substack{\vxdash : \vxdash_{\W} = \vxW \\ \vxdash\in\typ(\vX)}} \pact{\vxdash}.
\end{equation}
Rearranging the terms and using Taylor series expansions for $(1-\eps)^{-1}$ and $(1+\eps)^{-1}$, we have, with high probability, 
\begin{equation}
\label{eq:bounds-2}
(1 - \epsone)\left(1 - \eps + \Oh{\eps^2}\right)\phatm{\vxW} \leq \sum_{\substack{\vxdash : \vxdash_{\W} = \vxW \\ \vxdash\in\typ(\vX)}} \pact{\vxdash} \leq (1 + \epsone)\left(1 + \eps + \Oh{\eps^2}\right)\phatm{\vxW}.
\end{equation}

Now, consider the marginal innocent distribution $\pinnm{\vxW}$ as follows.
\begin{IEEEeqnarray}{rCl}
\pinnm{\vxW} & \stackrel{(i)}{=} & \pactm{\vxW} \\
& = & \sum_{\vxdash_{\Wc}} \pact{\vxdash_{\Wc},\vxW} \\
& = & \sum_{\vxdash:\vxdash_{\W} = \vxW} \pact{\vxdash} \\
\label{eq:sum-split}
& \stackrel{(j)}{=} & \sum_{\substack{\vxdash\in\typ(\vX) \\ \vxdash_{\W} = \vxW}} \pact{\vxdash} + \sum_{\substack{\vxdash\notin\typ(\vX) \\ \vxdash_{\W} = \vxW}} \pact{\vxdash},
\end{IEEEeqnarray}    
where (i) follows from since uses~\eqref{eq:convex-opti} and thus $\pactscal{.} = \pinnscal{.}$, and (j) follows by splitting the summation over typical and non-typical sets.

Finally, consider the total variation distance between the induced and the innocent marginal distributions
\begin{IEEEeqnarray}{rCl}
\Vdist{\phatm{.}}{\pinnm{.}} & = & \frac{1}{2}\sum_{\vxW}\left|\phatm{\vxW} - \pinnm{\vxW}\right| \\
& \stackrel{(k)}{=} & \frac{1}{2}\sum_{\vxW}\left|\phatm{\vxW} - \sum_{\substack{\vxdash\in\typ(\vX) \\ \vxdash_{\W} = \vxW}} \pact{\vxdash} + \sum_{\substack{\vxdash\notin\typ(\vX) \\ \vxdash_{\W} = \vxW}}\pact{\vxdash} \right| \IEEEeqnarraynumspace\\
& \leq & \frac{1}{2} \sum_{\vxW}\left| \phatm{\vxW} - \sum_{\substack{\vxdash\in\typ(\vX) \\ \vxdash_{\W} = \vxW}}\pact{\vxdash} \right| + \frac{1}{2}\sum_{\vxW} \sum_{\substack{\vxdash\notin\typ(\vX) \\ \vxdash_{\W} = \vxW}}\pact{\vxdash} \\
&\stackrel{(l)}{\leq} & \frac{1}{2} \sum_{\vxW} (\eps + \epsone)\phatm{\vxW} + \frac{1}{2}\sum_{\vxW} \sum_{\substack{\vxdash\notin\typ(\vX) \\ \vxdash_{\W} = \vxW}}\pact{\vxdash} \qquad\textrm{w.h.p.} \\
& \stackrel{(m)}{=} & \frac{1}{2}(\eps + \epsone) + \frac{1}{2} \sum_{\vxdash\notin\typ(\vX)}\pact{\vxdash} \\
& \stackrel{(n)}{\leq} & \frac{1}{2}(\eps + \epsone) + 2^{\C}\exp\left(-2\eps^2n\right)\\
& \stackrel{(o)}{=} & \eps + (2-\eps)2^{\C}\exp\left(-2\eps^2 n\right),
\end{IEEEeqnarray}
where (k) is due to~\eqref{eq:sum-split}, (l) is from~\eqref{eq:bounds-2} (neglecting the higher order terms), (m) is due to $\bigcup_{\vxW}\typ(\vXWc | \vxW) = \typ(\vX)$, (n) is due to typicality -- in particular, $\Pr{}_{\vX}\left(\vx\notin\typ\right) \leq 2\left|\calX\right|\exp{\left(-2\eps^2 n\right)}$ (see \cite{CsisK-book:2011}), where $\left|\calX\right|\ = 2^{\C}$, and (o) follows since $\epsone = \eps + (1-\eps)2^{\C + 1}\exp\left(-2\eps^2 n\right)$.
Therefore, for sufficiently large $n$, the total variation distance between the marginal innocent distribution and the marginal induced distribution can be made arbitrarily small with probability double exponentially close to $1$ by choosing $\eps = \Oh{\frac{1}{\sqrt{n}}}$, which completes the proof for deniability.

\textbf{Reliability :} Note that the encoding maps each $\vx\in\code$ with exactly one message $\m$. Thus, when Alice is active, Bob can always find $\tilde{m} : \vx\in\Bin{\tilde{m}}$, as the links are noiseless. When Alice is innocent, Bob's estimation fails if her innocent codeword falls in her active codebook. 

First, let us consider a case wherein the scalar active distribution differs from the scalar innocent distribution by more than $\eps$ for each symbol, {\it i.e.}, $\left|\pactscal(b) - \pinnscal{b}\right| \geq \eps$ $\forall\: b\in\calX$. In this case, the $\eps$-strongly typical set $\typ\left(\pactscal{.}\right)$ will have zero intersection with the $\eps$-strongly typical set $\typ\left(\pinnscal{.}\right)$. Since Alice's (active) codebook is a subset of $\typ\left(\pactscal{.}\right)$, the probability that an innocent codeword falls in her codebook is at the most the probability that the codeword is atypical, which itself is bounded below $2\left|\calX\right|\exp{\left(-2\eps^2 n\right)}$ (see \cite{CsisK-book:2011}). 

Next, consider the case where the scalar active and innocent distributions are close to each other. In this case, the worst scenario is when the given innocent distribution is such that $\pinnscal{.} = \pactscal{.}$. Here, Alice generates her codebook according to the distribution $\pact{.} = \pinn{.}$. When Alice is innocent, with high probability her innocent codeword $\vx\in\typ\left(\pinnscal{.}\right) = \typ\left(\pactscal{.}\right)$. Bob will wrongly estimate Alice to be active if her innocent codeword falls in her active codebook, which is equivalent to her innocent codeword falling in the first $\frac{2^{n\R}}{n}$ message bins. Due to random binning, the probability of this event is $\frac{1}{n}$. Alice's atypical codeword can never fall in her active codebook, since her codebook is a subset of $\eps$-strongly typical set. Therefore, Bob's estimation error probability is $\Oh{\frac{1}{n}}$.
When Alice is active, Bob can always decode the correct covert message, since the encoding is uniquely decodable. Thus, probability of decoding error is zero. Hence, the scheme is reliable.

\begin{remark}
\label{rem:corner-point-distributions}
Note that there exist some {\it corner point cases} in which $\pactscal{.}$ will be such that Willie can estimate the (sub)-codeword that he cannot observe. For example, consider a $3$-path network on which Willie can observe any two links. Suppose scalar active distribution is as follows. $\pactscal{\x} = \frac{1}{2}$ if $\x = 0$ or $\x = 7$, and $\pactscal{\x} = 0$ otherwise. In this case, solving the optimization results in $\pactscal{.} = \pinnscal{.}$ and further $\Hcond{\xWc}{\xW} = 0$. Therefore, any scheme cannot be deniable nor hidable, as Willie and Bob have equal power.
\end{remark}

\section{Deniable and Hidable Encoders}
\label{sec:hidable-encoders}
While deniability ensures that Willie cannot infer Alice's transmission status, it does not guarantee the hidability. Even if Willie is unable to estimate the transmission status, he can try to infer some information about the potential messages. At the same time, the hidability does not ensure deniability, since the notion of hidability focusses on securing the covert messages but not the fact that covert communication is under progress. We present some examples of the protocols that are deniable but not hidable, and vice-versa in the appendix~\ref{sec:deniability-vs-hidability}.

\subsection{Main Result for Deniable and Hidable Communication}
\label{sec:hidable-main-result}
First, we briefly review the concepts of {\it conditional typicality}, which we use in the proof of hidability (for details, see~\cite{CsisK-book:2011}). 

The {\it conditional type} of a codeword sequence $\vxWc$ given $\vxW$ is a stochastic matrix that gives the proportion of times a particular symbol of $\calXWc$ has occurred with each symbol of $\calXW$ in the pair $[\vxW; \vxWc]$. In particular, for $(a,b)\in \calXW \times \calXWc$, the conditional type is defined as $\typeveccond\left(b | a\right) = \frac{\N{a,b}{\vxW,\vxWc}}{\N{a}{\vxW}}$.

An $\eps$-conditionally strongly typical set is defined as
\begin{equation}
\label{eq:cond-typ-set-C-ary-1}
\typcond{\vXWc | \vxW} = 
\left\{\vxWc : 
\begin{array}{l} 
\left| \frac{1}{n} \N{a,b}{\vxW,\vxWc} - \pactscalcond{b}{a}\N{a}{\vxW} \right| \leq \frac{\eps}{\left|\calXW\right|} \\
\forall (a,b) \in \calXW \times \calXWc, \qquad \textrm{if} \quad \pactscalcond{b}{a} > 0\\
\N{a,b}{\vxW,\vxWc}  = 0, \qquad \textrm{if}  \quad \pactscalcond{b}{a} = 0
\end{array}
\right\}.
\end{equation}

Next, in the following, we show that the requirement of hidability in addition to deniability further reduces the rate at which Alice can communicate covert messages.  

\begin{theorem}
\label{thm:hidability-multilink}
The capacity of the {\it reliable-deniable-hidable} communication over a multipath network is
\begin{equation}
\label{eq:deniable-rate}
\Caphd = \sup_{\substack{\pscal{.} : \forall\W\in\bigW \\ \pscalW{.} = \pinnscalW{.}}} \min_{\W\in\bigW} \left[\Hp{\pscal{.}} - \Hp{\pscalW{.}}\right].
\end{equation}
In other words, for any sufficiently small $\delta$, any scalar innocent distribution $\pinnscal{.}$, any covert transmission rate $\Rhd < \Caphd$, there exists an encoder $\enc : \{0,1\}^{n\Rhd} \times \{0,1\}^{nr} \to \{0,1\}^{\C \times n}$ that is simultaneously $(1-\epsrel)$-reliable, $(1-\epsden)$-deniable, and $(1-\epshid)$-hidable for any $0 \leq \epsrel, \epsden, \epshid < \delta$ with high probability for sufficiently large block-length $n$. Conversely, any encoder with rate $\Rhd \geq \Caphd$ cannot be simultaneously hidable and deniable.
\end{theorem}

\subsection{Converse}
\label{sec:converse-hidability}
The converse follows from the standard information theoretic arguments as follows.
\begin{IEEEeqnarray}{rCl}
n\Rhd & \leq & \Hp{\M} \nonumber\\
& \stackrel{(a)}{\leq} & \Hcond{\M}{\vXW} + \epssec \nonumber\\
& = & \Hcond{\M}{\vXW} - \Hcond{\M}{\vX} + \Hcond{\M}{\vX} + \epssec, \nonumber\\
& \stackrel{(b)}{=} & \Hcond{\M}{\vXW} - \Hcond{\M}{\vXW,\vXWc} + n\epsn + \epssec, \nonumber\\
& = & \I{\M}{\vXWc | \vXW} + n\epsn + \epssec, \nonumber\\
& = & \Hcond{\vXWc}{\vXW} - \Hcond{\vXWc}{\M,\vXW}  + n\epsn+ \epssec, \nonumber\\
& \stackrel{(c)}{\leq} & \Hcond{\vXWc}{\vXW}  + n\epsn+ \epssec,\nonumber\\
\label{eq:hidable-rate}
& \stackrel{(d)}{\leq} & \min_{\W\in\bigW} \Hcond{\vXWc}{\vXW}  + n\epsn+ \epssec, \nonumber\\
& \stackrel{(e)}{\leq} & n \sup_{\substack{\pscal{.} : \forall\W\in\bigW \\ \pscalW{.} = \pinnscalW{.}}} \min_{\W\in\bigW} \Hcond{\XWc}{\XW}  + n\epsn+ \epssec \nonumber,
\end{IEEEeqnarray}
where $\epsn\to 0$ as $n\to\infty$, and $\epssec$ is some arbitrarily small positive constant. Here, (a) follows from the requirement of hidability (note that the hidability condition always guarantees the strong information-theoretic secrecy for some arbitrarily small $\epssec$), (b) follows from Fano's inequality and since $\vX = [\vXW,\vXWc]$, (c) follows from the non-negativity of entropy, (d) is due to Willie's ability to tap the subset of links having maximum entropy and (e) is due to independence bound of the entropy and the deniability condition.

\subsection{Achievability}
\label{sec:hidability-achievability}
\textbf{Encoding:} [Random Binning Based Stochastic Encoding] Codebook generation and encoding remains the same as in the deniability case, except that the rate used during the random binning stage is $\R = \min_{\W\in\bigW}\Hcond{\XWc}{\XW} - \Oh{\eps\log_2\frac{1}{\eps} + \frac{\log_2 n}{n}}$. The covert transmission rate is $\Rhd = \R - \frac{\log_2 n}{n} = \min_{\W\in\bigW}\Hcond{\XWc}{\XW} - \Oh{\eps\log_2\frac{1}{\eps} + \frac{\log_2 n}{n}} - \frac{\log_2 n}{n}$.

\textbf{Hidability:} First, we show that, for any sequence $\vxW$ observed by Willie, the number of codewords of a particular conditional type $\typeveccond\in\typcond{\vXWc|\vxW}$ per message bin is tightly concentrated around its mean value with high probability.
\begin{lemma}
\label{lem:concentrate-cond-type}
For any codeword $\vxW$ that is observed by Willie, for any $\typeveccond\in\typcond{\vXWc|\vxW}$, we have 
\begin{equation}
\label{eq:conc-cond-type}
\left| \typeveccond \cap\:\Bin{\m}\right| \to \expect{\left|\typeveccond\cap\:\Bin{\m}\right|} = \frac{\left|\typeveccond\right|}{2^{n\R}} \qquad \textrm{w.h.p.}
\end{equation}
provided
\begin{equation}
\label{eq:rate-hidable}
\R \leq \min_{\W\in\bigW}\Hp{\pactscalcond{.}{.}} - \constone, \quad \textrm{such that} \quad \constone \geq 2\eps \log_2 \frac{2^{\C}}{\eps} + \frac{2^{\C}\log_2(n+1)}{n}.
\end{equation}
\end{lemma}

To achieve hidability, we want to have (see~\eqref{eq:hidability}) 
$$\frac{Pr{}_{\vM,\enc}\left(\M = \m | \vx_\W, \T = 1\right)}{\Pr{}_{\vM}\left(\M = \m | \T = 1 \right)} \in \left[1-\epshid,1+\epshid\right].$$ 
Using Bayes' rule, the above condition transforms to
\begin{equation}
\label{eq:hidability-equivalent}
\frac{\Pr{}_{\vM,\enc}\left(\vX_{\W} = \vx_{\W} | \M = \m, \T = 1\right)}{\Pr{}_{\vM,\enc}\left(\vX_{\W} = \vx_{\W} | \T = 1\right)} \in \left[1-\epshid,1+\epshid\right].
\end{equation}

Now, since $\Pr{}_{\vM,\enc}\left(\vX_{\W} = \vx_{\W} | \T = 1\right) = \phatm{\vxW}$, by following the same lines as in the deniability proof, we can get ({\it cf.}~\eqref{eq:bounds-1}),
\begin{equation}
\label{eq:bounds-1-1}
\frac{1-\eps}{1+\epsone} \sum_{\substack{\vxdash : \vxdash_{\W} = \vxW \\ \vxdash\in\typ(\vX)}} \pact{\vxdash} \leq \Pr{}_{\vM,\enc}\left(\vX_{\W} = \vx_{\W} | \T = 1\right) \leq
\frac{1+\eps}{1-\epsone}\sum_{\substack{\vxdash : \vxdash_{\W} = \vxW \\ \vxdash\in\typ(\vX)}} \pact{\vxdash}.
\end{equation}

Next, consider the conditional probability of Willie observing a sequence $\vx_{\W}$ given that some message $\vm$ was transmitted as follows:
\begin{IEEEeqnarray}{rCl}
\IEEEeqnarraymulticol{3}{l}{\Pr{}_{\vM,\enc}\left(\vX_{\W} = \vx_{\W} | \vM = \vm, \T = 1\right)}\nonumber\\ 
\qquad \qquad & \stackrel{(a)}{=} & \sum_{\vxdash_{\Wc}} \Pr{}_{\vM,\enc}\left(\vxW, \vxdash_{\Wc} | \vM = \vm, \T = 1\right) \\
& = & \sum_{\vxdash : \vxdash_{\W}=\vx_{\W}} \Pr{}_{\vM,\enc}\left(\vxdash | \vM = \vm, \T = 1\right) \\
& = & \sum_{\vxdash : \vxdash_{\W}=\vx_{\W}} \pstochH{\vxdash}{\vm, \T=1} \\
& \stackrel{(b)}{=} & \sum_{\substack{\vxdash : \vxdash_{\W}=\vx_{\W} \\ \vxdash\in\Bin{\m}}} \frac{\pact{\vxdash}}{\Pr{}_{\mathcal{B}}\left(\Bin{\m}\right)},
\end{IEEEeqnarray}
where (a) follows from the rule of total probability and (b) follows from the definition of the stochastic encoder (see~\eqref{eq:stochastic-enc}). 

Using lemma~\ref{lem:bin-prob-bound} to bound $\Pr{}_{\mathcal{B}}\left(\Bin{\vm}\right)$\footnote{Notice that the rate used for random binning $\R = \min_{\W\in\bigW}\Hcond{\XWc}{\XW} - \Oh{\eps\log_2\frac{1}{\eps} + \frac{\log_2 n}{n}}$ is less than the bound given in~\eqref{eq:super-rate-deniability}, and thus, Lemma~\ref{lem:bin-prob-bound} can be used.}, and the isomorphism between the sets $\left\{\vxdash : \vxdash\in\typ, \vxdash_{\W}=\vxW\right\}$ and $\typcond{\vXWc | \vxW}$ which follows from Lemma~\ref{lem:Willie-typicality}, it is easy to show that
\begin{IEEEeqnarray}{rCl}
\IEEEeqnarraymulticol{3}{l}{\frac{2^{n\R}}{1+\epsone}\sum_{\typeveccond\in\typcond{\vXWc | \vxW}} \left|\typeveccond\cap\:\Bin{\vm}\right| \pact{\vxdash_{\typevec}} \leq \Pr{}_{\vM,\enc}\left(\vx_{\W} | \vm, \T = 1\right)} \nonumber\\
\IEEEeqnarraymulticol{3}{l}{\qquad\qquad \leq \frac{2^{n\R}}{1-\epsone}\sum_{\typeveccond\in\typcond{\vXWc | \vxW}} \left|\typeveccond\cap\:\Bin{\vm}\right| \pact{\vxdash_{\typevec}},}
\end{IEEEeqnarray}
where $\vxdash_{\typevec} = \{\vxdash : \vxdash_{\Wc} \in \typeveccond\}$. 

From Lemma~\ref{lem:concentrate-cond-type}, as condition~\eqref{eq:rate-hidable} is satisfied, we have, with high probability, that 
\begin{IEEEeqnarray}{rCl}
\IEEEeqnarraymulticol{3}{l}{2^{n\R} \frac{1-\eps}{1+\epsone}\sum_{\typeveccond\in\typcond{\vXWc | \vxW}} \frac{\left|\typeveccond\right|}{2^{n\R}} \pact{\vxdash_{\typevec}} \leq \Pr{}_{\vM,\enc}\left(\vx_{\W} | \vm, \T = 1\right)} \nonumber\\
\IEEEeqnarraymulticol{3}{l}{\qquad\qquad \leq 2^{n\R} \frac{1+\eps}{1-\epsone}\sum_{\typeveccond\in\typcond{\vXWc | \vxW}} \frac{\left|\typeveccond\right|}{2^{n\R}} \pact{\vxdash_{\typevec}}.}
\end{IEEEeqnarray}
However, since $\sum_{\typeveccond\in\typcond{\vXWc | \vxW}}\left|\typeveccond\right| \pact{\vxdash_{\typevec}} = \sum_{\vxdash:\vxdash_{\Wc}\in\typcond{\vXWc | \vxW}} \pact{\vxdash}$, above inequalities are equivalent to 
\begin{IEEEeqnarray}{rCl}
\IEEEeqnarraymulticol{3}{l}{\frac{1-\eps}{1+\epsone}\sum_{\vxdash:\vxdash_{\Wc}\in\typcond{\vXWc | \vxW}} \pact{\vxdash} \leq \Pr{}_{\vM,\enc}\left(\vx_{\W} | \vm, \T = 1\right)} \nonumber\\
\IEEEeqnarraymulticol{3}{l}{\qquad\qquad \leq \frac{1+\eps}{1-\epsone}\sum_{\vxdash:\vxdash_{\Wc}\in\typcond{\vXWc | \vxW}} \pact{\vxdash}.}
\end{IEEEeqnarray}
Finally, using the isomorphism between the sets $\left\{\vxdash : \vxdash\in\typ, \vxdash_{\W}=\vxW\right\}$ and $\typcond{\vXWc | \vxW}$ (Lemma~\ref{lem:Willie-typicality}), with high probability, we have
\begin{equation}
\label{eq:bounds-hid}
\frac{1-\eps}{1+\epsone}\sum_{\substack{\vxdash : \vxdash_{\W} = \vxW \\ \vxdash\in\typ(\vX)}}  \pact{\vxdash} \leq \Pr{}_{\vM,\enc}\left(\vx_{\W} | \vm, \T = 1\right) \leq \frac{1+\eps}{1-\epsone}\sum_{\substack{\vxdash : \vxdash_{\W} = \vxW \\ \vxdash\in\typ(\vX)}} \pact{\vxdash}.
\end{equation}

From~\eqref{eq:bounds-1-1} and~\eqref{eq:bounds-hid}, we have, with high probability,
\begin{equation}
\label{eq:hidable-bound}
\frac{(1-\eps)(1-\epsone)}{(1+\eps)(1+\epsone)} \leq \frac{\Pr{}_{\vM,\enc}\left(\vX_{\W} = \vx_{\W} | \vM = \vm, \T = 1\right)}{\Pr{}_{\vM,\enc}\left(\vX_{\W} = \vx_{\W} | \T = 1\right)} \leq \frac{(1+\eps)(1+\epsone)}{(1-\eps)(1-\epsone)},
\end{equation}
when the covert transmission rate is $\Rhd = \min_{\W\in\bigW}\Hp{\XWc|\XW} - \Oh{\eps\log_2{\frac{2^{\C}}{\eps}} + \frac{\log_2 n}{n}} - \frac{\log_2 n}{n}$, and hence the proof.

\textbf{Deniability and Reliability:} Since the only change from the deniability case is reduction in the total number of bins used in the binning step, analysis the the previous section holds and the scheme continues to be deniable and reliable. In particular, in section~\ref{sec:deniability-achievability}, we showed that the random binning based stochastic encoding is deniable for $\Rd = \Hp{\pinnscal{.}} - \Oh{\eps\log_2{\frac{2^{\C}}{\eps}} + \frac{2^{\C}\log_2(n+1)}{n}} - \frac{\log_2 n}{n}$. As conditioning reduces entropy, $\Rhd \leq \Rd$, and thus, previous analysis holds.

\begin{remark}
Conventional information-theoretically secure techniques, like mixing random keys, essentially exploit the stochastic nature of the encoding to achieve secrecy. Notice that the encoding used for deniability is inherently stochastic. We show that by appropriately reducing the total number of rate, it is possible to achieve hidability in addition to deniability. Intuitively, reducing the total number of bins increases the number of typical sequences per bin, which in turn enhances the {\it level of randomness}.
\end{remark}

\section{Conclusion}
In this paper, we characterized the capacity of for {\it reliable-deniable} communication over a multipath network and presented an achievability strategy based on random binning based stochastic encoding. Further, we proposed the concept of hidability, which is a stronger condition than strong information-theoretic secrecy. Finally, we characterized the capacity for {\it reliable-deniable-hidable} communication over multipath networks.


%

\appendices
\section{Hidability vs. Information Theoretic Secrecy}
\label{sec:hidability-vs-secrecy}
In this appendix, we compare the condition of hidability with the condition for strong information-theoretic secrecy. Recall that a scheme is said to be $(1-\epshid)$-hidable if, for some small $\epshid >0$, we have
\begin{equation}
\label{eq:hidability-1}
\frac{\Pr{}_{\vM,\enc}\left(\vM = \vm | \vxW, \T = 1\right)}{\Pr{}_{\vM}\left(\vM = \vm | \T = 1 \right)} \in \left[1 - \epshid , 1 + \epshid\right], \quad \forall \: \vm, \: \forall \: \W \in \bigW,
\end{equation}
for any $\left|\W\right| \times n$ binary matrix $\vxW$ that is observed by Willie. On the other hand, for any scheme to satisfy strong information-theoretic secrecy, we need to have \cite{MauW:00}:
\begin{equation} 
\label{eq:info-theoretic-secrecy}
\I{\vM}{\vXW | \T = 1} \leq \epssec, \qquad \forall \W\in\bigW,
\end{equation}
for some arbitrarily small $\epssec > 0$.

In the following, we show that the hidability conditional  is a stronger condition than the strong information-theoretic secrecy condition.
\begin{theorem}
\label{thm:hidability-vs-secrecy}
If a communication scheme is $\left(1 - \oh{\frac{1}{n}}\right)$-hidable, then it must have the strong information-theoretic security. The converse is not necessarily true.
\end{theorem}
\begin{IEEEproof}
Suppose that the communication scheme under consideration is $(1-\epshid)$-hidable. Consider the  entropy of the covert messages conditioned on Willie's particular observation given that Alice is in the active status, as  follows:
\begin{IEEEeqnarray}{rCl}
\Hcond{\vM}{\vXW = \vxW,\T=1} & = & - \sum_{\vm} \Pr{}_{\vM,\enc}\left(\vm | \vxW,\T=1\right) \log_2\left[{\Pr{}_{\vM,\enc}\left(\vm | \vxW,\T=1\right)}\right] \IEEEeqnarraynumspace\\
& \stackrel{(a)}{\geq} & - \sum_{\vm} \left(1 - \epshid \right) \Pr{}_{\vM}\left(\vm | \T = 1\right) \log_2\left[\left(1+\epshid\right)\Pr{}_{\vM}\left(\vm | \T = 1\right)\right] \\
\label{eq:LB-cond-entropy}
& = & \left(1 - \epshid\right) \Hcond{\vM}{\T=1} - \left(1 - \epshid\right)\log_2\left(1 + \epshid\right), 
\end{IEEEeqnarray}
where (a) follows from the assumption that the scheme is $(1-\epshid)$-hidable. Then, we have
\begin{IEEEeqnarray}{rCl}
\Hcond{\vM}{\vXW,\T=1} & = & \sum_{\vxW}\Pr{}_{\vM,\enc}\left(\vxW | \T=1\right) \Hcond{\vM}{\vXW = \vxW,\T=1} \\
& \stackrel{(b)}{\geq} & \sum_{\vxW} \Pr{}_{\vM,\enc}\left(\vxW | \T=1\right) \left(1 - \epshid\right)\left[\Hcond{\vM}{\T=1} - \log_2\left(1 + \epshid\right)\right] \IEEEeqnarraynumspace\\
\label{eq:LB-cond-entropy-1}
& = & \left(1 - \epshid\right) \left[\Hcond{\vM}{\T=1} - \log_2\left(1 + \epshid\right)\right],
\end{IEEEeqnarray}
where (b) follows from the lower bound~\eqref{eq:LB-cond-entropy}. Therefore, we can write
\begin{IEEEeqnarray}{rCl}
\I{\vM}{\vXW | \T = 1} & = & \Hcond{\vM}{\T=1} - \Hcond{\vM}{\vXW,\T=1} \\
& \stackrel{(c)}{\leq} & \Hcond{\vM}{\T=1} - \left(1 - \epshid\right) \left[\Hcond{\vM}{\T=1} - \log_2\left(1 + \epshid\right)\right] \\
& = & \epshid \Hcond{\vM}{\T=1} + \left(1 - \epshid\right)\log_2\left(1 + \epshid\right) \\
& \stackrel{(d)}{\leq} & \epshid n\Rs + \left(1 - \epshid\right)\log_2\left(1 + \epshid\right),
\end{IEEEeqnarray}
where (c) follows from~\eqref{eq:LB-cond-entropy-1}, and (d) follows because $\Hcond{\vM}{\T=1} \leq \log_2\left|\vM\right| = {n\Rs}$, where $\Rs$ is the rate of transmission for the covert messages. Notice that any of the arguments above does not depend upon the specific choice of $\W$. Hence, $\forall \W\in\bigW$,  we have $\I{\vM}{\vXW | \T = 1} \leq \epshid n\Rs + \Oh{\epshid}$, and thus, if we choose $\epshid = \frac{\epssec}{n\Rs}$, we will achieve information-theoretic secrecy. Moreover, if $\epshid = \oh{\frac{1}{n}}$, the hidability implies the strong information-theoretic security for any arbitrarily small $\epssec$.

To show that the converse may not necessarily be true, we give an example of a strongly secure scheme that is not hidable. Even though we do not worry about the deniability in this example, it is possible to extend the presented approach to deniable schemes. Consider a $\C = 2$ link multipath network, with Willie observing any one of the links. In the active state, Alice employees a slight variant of Shannon's one-time padding scheme as follows. 

Alice, first, randomly selects a subset $\mathcal{D}$ of binary, length-$n$ sequences of cardinality $\left|\mathcal{D}\right| = 2^{n\delta}$ for some small $\delta \geq 0$. While transmitting a particular covert message sequence $\vm$, if $\vm\notin\mathcal{D}$, Alice uses Shannon's one-time padding. Specifically, she generates a binary, length-$n$ key sequence $\vk$ uniformly at random from the set $\mathcal{D}^c$. Then, she transmits the key sequence $\vk$ on the top link, and the exclusive OR of the key sequence and the covert message on the bottom link. However, if $\vm\in\mathcal{D}$, she simply transmits the message $\vm$ on the top as well as the bottom link.  

First, let us show that this scheme is strongly secure. Notice that, since Willie knows the encoding scheme, he knows the set $\mathcal{D}$. Without loss of generality, assume that Willie is observing the top link. If Willie observes a sequence from the set $\mathcal{D}$, then he knows that Alice is transmitting that particular message. Hence, we have
\begin{equation}
\label{eq:cond-entropy-ex}
\Hcond{\vM}{\vXW = \vxW} = 
\begin{cases}
0 & \textrm{if} \quad \vxW \in \mathcal{D} \\
\log_2\left(2^n - 2^{n\delta}\right) & \textrm{if} \quad \vxW \notin \mathcal{D}
\end{cases}
\end{equation}
Therefore, we have 
\begin{equation}
\label{eq:cond-entropy-ex-1}
\Hcond{\vM}{\vXW} = \left(1 - \frac{2^{n\delta}}{2^n}\right)\log_2\left(2^n - 2^{n\epsilon}\right).
\end{equation}
Then, it is easy to show that
\begin{IEEEeqnarray}{rCl}
\label{eq:mutual-info-ex}
\I{\vM}{\vXW} & = & H_2\left(\frac{2^{n\delta}}{2^n}\right) + \frac{2^{n\delta}}{2^n}\log_2\left(2^{n\delta}\right),  \\
& = & \oh{1} \qquad \textrm{for} \: n \rightarrow \infty,
\end{IEEEeqnarray}
where $H_2(.)$ denotes the binary entropy function.

Next, we show that this scheme does not satisfy the hidability criterion. For instance, for any sequence $\vm_{\mathcal{D}} \in \mathcal{D}$, we will have $\Pr{}_{\vM,\enc}\left(\vM = \vm_{\mathcal{D}} | \vXW = \vm_{\mathcal{D}}, \T = 1\right) = $, and thus,
$$\frac{\Pr{}_{\vM,\enc}\left(\vM = \vm | \vXW = \vm, \T = 1\right)}{\Pr{}_{\vM}\left(\vM = \vm | \T = 1\right)} = 2^n \qquad \forall \vm\in\mathcal{D}, \: \forall\W\in\bigW,$$ 
violating the hidability. Intuitively, if Willie happens to observe a sequence from the set $\mathcal{D}$, he knows which message was transmitted, and the messages in the set $\mathcal{D}$ are not secure. One can argue that such feeble  messages are asymptotically small in fraction. However, one should note that there cardinality is symptomatically very large. Moreover, looking at this from the reverse perspective clearly depicts the weakness of this non-hidable encoding scheme, as follows.

Suppose Willie is specifically interested in whether any of the message in set $\mathcal{D}$ is transmitted or not. Then, when he observes any sequence $\vxW\notin\mathcal{D}$, he knows that no message in the set $\mathcal{D}$ was transmitted.  Notice that such event occurs with a large probability of $\left(1 - \frac{2^{n\delta}}{2^n}\right)$.

\end{IEEEproof}

\section{Examples on Deniability vs. Hidability}
\label{sec:deniability-vs-hidability}
In this section, we present some examples of the schemes that are hidable but not deniable and vice-versa.

First, let us consider a hidable scheme that is not deniable. For this, consider Shannon's one-time padding scheme described in the toy example in section~\ref{sec:problem} (see Fig.~\ref{fig:ex-1} on page 6). It is straightforward to show that this scheme is hidable. However, it induces uniform distribution on the codewords on each link. The marginal induced distribution is far away from the marginal innocent distribution, and thus, the scheme is not deniable. In fact, notice that this scheme will be deniable only if the marginal innocent distributions on both the individual links are close to uniform.

\begin{figure}[!t]
 \centering
  \subfigure[System Diagram]{
  \label{fig:den-not-hid-a}
  \def\svgwidth{500pt}
  \small{
  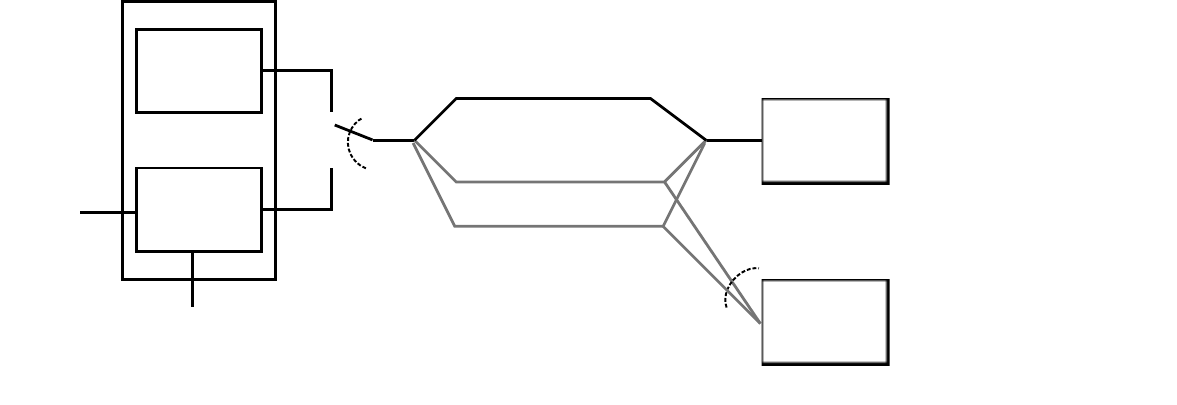
  }
  } \\
  \subfigure[Scalar Innocent Distribution]{
  \label{fig:den-not-hid-b}
  \begin{tabular}{|c|c|}
  \hline
  $\X(t) = \left[\X_1(t)\: \X_2(t) \: \X_3(t)\right]^T$ & $\pinnscal{.}$ \\
  \hline
  $0\: 0\: 0$ & $\frac{1}{2}$ \\
  $0\: 0\: 1$ & 0\\
  $0\: 1\: 0$ & 0\\
  $0\: 1\: 1$ & 0\\
  $1\: 0\: 0$ & 0 \\
  $1\: 0\: 1$ & 0\\
  $1\: 1\: 0$ & 0\\
  $1\: 1\: 1$ & $\frac{1}{2}$\\
  \hline
  \end{tabular}} \qquad \qquad
  \subfigure[Scalar Marginal Innocent Distribution on any pair of links $\{j,k\}$]{
  \label{fig:den-not-hid-c}
  \begin{tabular}{|c|c|}
  \hline
  $\X(t) = \left[\X_j(t)\: \X_k(t)\right]^T$ & $\pinnscalW{.}$ \\
  \hline
  $0\: 0$ & $\frac{1}{2}$\\
  $0\: 1$ & 0\\
  $1\: 0$ & 0\\
  $1\: 1$ & $\frac{1}{2}$\\
  \hline
  \end{tabular}}
  \caption{Example of a multipath network with $\C = 3$ links. Willie can observe any pair of the links.} 	
  \label{fig:den-not-hid}
\end{figure}

Now, consider an example of a deniable scheme that is not hidable. Consider a multipath network with $\C = 3$ links. Willie can observe any pair of links. The scalar innocent distribution is given by Fig.~\ref{fig:den-not-hid-b}. This results in the same marginal innocent distribution as given in Fig.~\ref{fig:den-not-hid-c} on each pair of links. In this case, Alice can use a simple scheme in the active state: she simply copies her cover message bit on each link. In this way, Alice can transmit one bit of covert message to Bob per network use. Since the covert message bits are assumed to be uniformly distributed, this scheme is deniable. 

However, this scheme is clearly not hidable.  This is because Willie can perfectly decode Alice's message from his observations, when he knows that Alice is in the active status. This makes the ratio $\frac{\Pr{}_{\vM,\enc}\left({\vMW} = \vm | \vxW, \T = 1\right)}{\Pr{}_{\vM}\left({\vMW} = \vm | \T = 1 \right)} = 2^n$, which is exponentially large.

\section{Proofs of Lemmas}
\label{sec:useful-lemmas}

\subsection{Proof of Lemma~\ref{lem:concentration-typeclass}}
Note that $\left\vert{\typevec \cap \Bin{\m}}\right\vert$ denotes the number of codewords of type class $\typevec$ that are {\it fallen} in the bin $\Bin{\m}$. For any particular type class $\typevec$, we have $|\typevec|$ codewords which will be {\it thrown} uniformly at random in $2^{n\R}$ message bins. Therefore, the expected number of codewords of a particular type class which fall into a given message bin is $\expect{\left|\typevec\cap\Bin{\vm}\right|} = \frac{\left|\typevec\right|}{2^{n\R}}$.
Moreover, note that for any type $\typevec$ defined over an alphabet $\calX$, the type class $\typevec$ satisfies (see \cite{CovT:12})
\begin{equation}
\label{eq:typeclass-cardinality-bounds}
\frac{1}{(n+1)^{\left|\calX\right|}}2^{n\Hp{\typevec}} \leq \left| \typevec \right| \leq 2^{n\Hp{\typevec}}
\end{equation}
Hence, using $\left|\calX\right| = 2^{\C}$, we get
\begin{equation}
\label{eq:typeclass-exp-cardinality-bounds}
\frac{1}{(n+1)^{2^{\C}}}\frac{2^{n\Hp{\typevec}}}{2^{n\R}} \leq \expect{\left| \typevec \cap \Bin{\vm}\right|} \leq \frac{2^{n\Hp{\typevec}}}{2^{n\R}}.
\end{equation} 
Then, by using Chernoff bound, we can write
\begin{IEEEeqnarray}{rCl}
\IEEEeqnarraymulticol{3}{l}{\Pr\left(\Big| {\left\vert \typevec\cap \Bin{\vm}\right\vert - \expect{\left\vert \typevec \cap \Bin{\vm}\right\vert}} \Big| \geq \epsilon \: \expect{\left\vert \typevec \cap \Bin{\vm}\right\vert}\right)} \nonumber \\ 
\qquad & \leq &
2\exp{\left(\frac{-\epsilon^2 \expect{\left|{\typevec\cap\Bin{\vm}}\right|}}{3}\right)}, \\
\label{eq:concentration-upper-bound}
& \stackrel{(a)}{\leq} & 2\exp{\left(\frac{-\epsilon^2 2^{n\Hp{\typevec}}}{3(n+1)^{2^{\C}}2^{n\R}}\right)}, 
\end{IEEEeqnarray}
where (a) follows from the lower bound in~\eqref{eq:typeclass-exp-cardinality-bounds}. 

After simplifying~\eqref{eq:concentration-upper-bound}, we can see that, if $\R = \Hp{\pactscal{.}} - \const$, then as $n\to\infty$, we have $\left|\typevec\cap\Bin{\vm}\right| \to \expect{\left|\typevec\cap\Bin{\vm}\right|}$ with probability at least $1 - 2 \exp{\left(\frac{-\eps^2 2^{n\left[\Hp{\typevec} - \frac{2^{\C}\log_2(n+1)}{n} - \Hp{\pactscal{.}} + \const\right]}}{3}\right)}$.
Note that this probability will approach 1 if the exponent term $\left[\Hp{\typevec} - \frac{2^{\C}\log_2(n+1)}{n} - \Hp{\pactscal{.}} + \const\right]$ is positive $\forall{\typevec\in\typ(\vX)}$. For this, it suffices to consider $\typevec^* = \arg\min_{\typevec\in\typ(\vX)}\Hp{\typevec}$, which is the type class in the typical set such that the entropy of the corresponding type is the minimum. This type $\typevec^{*}$ will enable us to characterize the value of $\const$ such that the exponent term is positive $\forall{\typevec\in\typ(\vX)}$. 

Let $\theta(\typevec^{*})$ denote the total variation distance between type $\typevec^{*}$ and $\pactscal{.}$. Then, we have
\begin{IEEEeqnarray}{rCl}
\theta(\typevec^{*}) = \Vdist{\typevec^{*}}{\pinnscal{.}} & = & \frac{1}{2}\sum_{a\in\calX} \left| \typevec^{*}(a) - \pactscal{a}\right|, \\
\label{eq:by-strong-typicality}
& \leq & \frac{1}{2}\sum_{a\in\calX} \frac{\eps}{\left|\calX\right|}, \\
& = & \frac{\eps}{2},
\end{IEEEeqnarray}
where equation~\eqref{eq:by-strong-typicality} follows from the definition of strongly typical set. Now, for any $\eps \leq \frac{1}{2}$, we will have $\theta(\typevec^{*}) \leq \frac{1}{4}$, which would allow us to use the following result from \cite[Lemma 2.7]{CsisK-book:2011}:
\begin{equation}
\label{eq:bound-diff-in-entropies}
\left| \Hp{\typevec^{*}} - \Hp{\pactscal{.}} \right| \leq - 2\theta(\typevec^{*})\log_2 \frac{2\theta(\typevec^{*})}{\left|\calX\right|}.
\end{equation}
Moreover, since the RHS in~\eqref{eq:bound-diff-in-entropies} is an increasing function of $\theta(\typevec^{*})$ for $0\leq\theta(\typevec^{*})\leq\frac{1}{2}$, we have $\left| \Hp{\typevec^{*}} - \Hp{\pactscal{.}} \right| \leq -\eps\log_2\frac{\eps}{\left|\calX\right|}$. Therefore, we get $\left[\Hp{\typevec^*} - \frac{2^{\C}\log_2(n+1)}{n} - \Hp{\pactscal{.}} + \const \right] \geq \eps\log_2\frac{\eps}{\left|\calX\right|} - \frac{2^{\C}\log_2(n+1)}{n} + \const$, which is positive if~\eqref{eq:super-rate-deniability} holds. Consequently, $\left|\typevec\cap\Bin{\vm}\right| \to \expect{\left|\typevec\cap\Bin{\vm}\right|}$ with probability approaching double exponentially close to 1 if~\eqref{eq:super-rate-deniability} holds.

\subsection{Proof of Lemma~\ref{lem:bin-prob-bound}}
\label{sec:proof-2}
Beginning with the definition of the probability of a message bin, we can write
\begin{IEEEeqnarray}{rCl}
\label{eq:bin-prob}         
\Pr{}_{\mathcal{B}}\left(\Bin{\vm}\right) & = & \sum_{\vx : \vx\in\code\cap\Bin{\vm}}\pact{\vx}, \\
& \stackrel{(a)}{=} & \sum_{\vx \in \typ \cap \Bin{\vm}} \pact{\vx}, \\
\label{eq:break-into-typeclass}
& \stackrel{(b)}{=} & \sum_{\typevec \in \typ} \sum_{\vx \in \typevec \cap \Bin{\vm}} \pact{\vx} \\
\label{eq:prob-type-class-in-card}
& \stackrel{(c)}{=} & \sum_{\typevec \in \typ} \left\vert{\typevec \cap \Bin{\vm}}\right\vert \pact{\vx : \vx\in\typevec},
\end{IEEEeqnarray}  
where (a) follows since in the random binning based schemes, the codebook is the $\eps$-strongly typical set, {\it i.e.}, $\code = \typ$, (b) follows because the strongly typical set can be considered as a collection of several type classes, and (c) follows from the fact that the probability of each codeword in any particular type class is the same. Notice that, here, $\left\vert{\typevec \cap \Bin{\vm}}\right\vert$ denote the number of codewords of $\typevec$ that are associated with the bin $\Bin{\vm}$. 
                     
Taking expectation of both sides with respect to the binning process, we get
\begin{IEEEeqnarray}{rCl}                                    
\label{eq:exp-bin-prob}
\expect{\Pr{}_{\mathcal{B}}\left(\Bin{\vm}\right)} & = & \sum_{\typevec \in \typ} 
\expect{\left\vert{\typevec \cap \Bin{\vm}}\right\vert} \pact{\vx : \vx\in\typevec} \\
\label{eq:exp-balls-in-bin}
& \stackrel{(d)}{=} & \sum_{\typevec \in \typ} 
\frac{\left|\typevec\right|}{2^{n\R}} \pact{\vx : \vx\in\typevec} \\
\label{eq:prob-typeclass-sum}
& = & \frac{1}{2^{n\R}}\sum_{\typevec \in \typ} \left|\typevec\right| \pact{\vx : \vx\in\typeclass_j} \\
\label{eq:prob-typ-set}
& \stackrel{(e)}{=} & \frac{1}{2^{n\R}} \Pr{}_{\vX}\left(\typ\left(\pactscal{.}\right)\right) \\
\end{IEEEeqnarray}
where, to get (d), recall that in a random binning experiment with $m$ balls and $k$ bins, the expected number of balls in a bin is $\frac{m}{k}$. For (e), observe that $\left|\typevec\right| \pact{\vx : \vx\in\typevec} = \Pr{}_{}\left(\typevec\right)$. Finally, using the the probability of $\eps$-strongly typical set can be lower bounded as $\Pr\left(\typ\right) \geq 1-2\left|\calX\right|\exp{\left(- 2 \eps^2 n\right)}$ (see \cite{CsisK-book:2011}). Therefore, we have 
\begin{equation}
\label{eq:range-exp-bin-prob}
\expect{\Pr\left(\Bin{\vm}\right)} \in \left(1-2^{\C+1}\exp{\left(-2\eps^2 n\right)},1\right)\frac{1}{2^{n\R}},
\end{equation}
where we substitute $\left|\calX\right| = 2^{\C}$.

Now, consider
\begin{IEEEeqnarray}{rCl}
\label{eq:prob-diff}
\left|{\Pr{}_{\mathcal{B}}{\left(\Bin{\vm}\right)} - \expect{\Pr{}_{\mathcal{B}}\left(\Bin{\vm}\right)}}\right| & \stackrel{(g)}{=} & \left\vert
\sum_{\typevec \in \typ} \left|\typevec \cap \Bin{\vm}\right| \pact{\vx : \vx\in\typevec} \right. \\
&  & \: -  
\left. \sum_{\typevec \in \typ} \expect{\left|\typevec \cap \Bin{\vm}\right|} \pact{\vx : \vx\in\typevec} \right\vert, \\
& \leq & \sum_{\typevec \in \typ} \Big|{\left[{\left|\typevec \cap \Bin{\vm}\right| - \expect{\left|\typevec \cap \Bin{\vm}}\right|}\right]}\Big|  \pact{\vx : \vx\in\typevec}, \IEEEeqnarraynumspace\\
\label{eq:concentration-1}
& \stackrel{(h)}{\leq} & \sum_{\typevec \in \typ} \eps \: \expect{\left\vert{\typevec \cap \Bin{\vm}}\right\vert} \pact{\vx : \vx\in\typevec}, \qquad \textrm{w.h.p.}\\
& \stackrel{(i)}{=} & \eps \: \expect{\Pr{}_{\mathcal{B}}{\left(\Bin{\vm}\right)}}, \\
\label{eq:prob-bin}
\therefore \quad \Pr{}_{\mathcal{B}}{\left(\Bin{\vm}\right)} & \in & \left[(1 - \eps), (1+\eps)\right]\expect{\Pr{}_{\mathcal{B}}{\left(\Bin{\vm}\right)}}, \qquad \textrm{w.h.p.}
\end{IEEEeqnarray}     
where (g) follows from~\eqref{eq:prob-type-class-in-card}, (h) follows from the result proven in the first half, and (i) follows again from~\eqref{eq:prob-type-class-in-card}. 
Finally, the result~\eqref{eq:bin-prob-bound} follows from~\eqref{eq:range-exp-bin-prob} and~\eqref{eq:prob-bin}.

\subsection{Proof of Lemma~\ref{lem:Willie-typicality}}
\label{sec:proof-3}
Note that the transmitted codeword can be decomposed into the part observed by Willie and the part not observed by Willie as $\vx = \left[\vxW ; \: \vxWc\right]$, 
where $\vx \in \calX^n = {\left\{0,1,\ldots,2^{\C}-1\right\}}^n$, $\vxW \in \calXW^n = {\left\{0,1,\ldots,2^{\left|\W\right|}-1\right\}}^n$, and $\vxWc \in \calXWc^n = {\left\{0,1,\ldots,2^{\left|\Wc\right|}-1\right\}}^n$. Further, each symbol $c\in\calX$ of the codeword is associated with a pair of symbols $(a,b)\in\calXW \times \calXWc$. 

Now, for a strongly typical codeword $\vx\in\typ$, we have
\begin{IEEEeqnarray}{rCl}
\left|\frac{1}{n}\N{c}{\vx} - \pactscal{c}\right| &\leq & \frac{\eps}{\left|\calX\right|}\qquad\forall c\in\calX \nonumber\\
\label{eq:jointly-typical}
\therefore \quad \left|\frac{1}{n}\N{a,b}{\vxW,\vxWc} - \pactscalWWc{a,b}\right| & \leq & \frac{\eps}{\left|\calXW\right|\left|\calXWc\right|} \quad\forall (a,b)\in\calXW \times \calXWc,
\end{IEEEeqnarray}
where distribution $\pactscal{.}$ can be represented as the joint distribution $\pactscalWWc{.,.}$.
Then, for a sequence $\vxW$ observed by Willie, we have
\begin{IEEEeqnarray}{rCl}
\left|\frac{1}{n}\N{a}{\vxW} - \pactscalW{a}\right| & = & \left|\sum_{b\in\calXWc}\frac{1}{n}\N{a,b}{\vxW,\vxWc} - \sum_{b\in\calXWc}\pactscal{a,b}\right| \qquad\forall a\in\calXW \nonumber\\
& \leq & \sum_{{b\in\calXWc}}\left|\frac{1}{n}\N{a,b}{\vxW,\vxWc} - \pactscalWWc{a,b}\right|\qquad\forall a\in\calXW\nonumber\\
\label{eq:Willie-typical}
& \stackrel{(p)}{\leq} & \frac{\eps}{\left|\calXW\right|}\qquad\forall a\in\calXW,
\end{IEEEeqnarray}
where (p) follows from the strong typicality of $\vx$ (see~\eqref{eq:jointly-typical}). Hence, when $\vx$ is strongly typical w.r.t. $\pactscal{.}$, $\vxW$ is strongly typical w.r.t. $\pactscalW{.}$.

Now, in order to prove the conditional strong typicality of $\vxWc$ given $\vxW$, observe that
\begin{IEEEeqnarray}{rCl}
\IEEEeqnarraymulticol{3}{l}{\frac{1}{n}\left|\N{a,b}{\vxW,\vxWc} - \pactscalcond{b}{a}\N{a}{\vxW}\right|} \nonumber\\
\qquad & \stackrel{(q)}{\leq} & \frac{\eps}{\left|\calXW\right|}\left(\frac{1}{\left|\calXWc\right|}  + \pactscalcond{b}{a}\right) \qquad\forall(a,b)\in\calXW \times \calXWc,\\
\label{eq:Willie-cond-typical}
& \leq & \frac{\eps}{\left|\calXW\right|}\left(\frac{1+\left|\calXWc\right|}{\left|\calXWc\right|}\right),
\end{IEEEeqnarray} 
where (q) follows from~\eqref{eq:jointly-typical} and~\eqref{eq:Willie-typical}, thus, proving the conditional strong typicality of $\vxWc$ given $\vxW$. Note that equation~\eqref{eq:Willie-cond-typical} imposes the definition of the conditionally strongly typical set as
\begin{equation}
\label{eq:cond-typ-set-C-ary-1}
\typcond{\vXWc | \vxW} = 
\left\{\vxWc : 
\begin{array}{l} 
\left| \frac{1}{n} \N{a,b}{\vxW,\vxWc} - \pactscalcond{b}{a}\N{a}{\vxW} \right| \leq \frac{\eps\left(1+\left|\calXWc\right|\right)}{\left|\calXW\right| \left|\calXWc\right|} \\
\forall (a,b) \in \calXW \times \calXWc, \qquad \textrm{if} \quad \pactscalcond{b}{a} > 0\\
\N{a,b}{\vxW,\vxWc}  = 0, \qquad \textrm{if}  \quad \pactscalcond{b}{a} = 0
\end{array}
\right\}.
\end{equation}

\subsection{Proof of Lemma~\ref{lem:concentrate-cond-type}}
\label{sec:proof-4}
Recall that the conditional type of a codeword sequence $\vxWc$ given $\vxW$ is a stochastic matrix that gives the proportion of times a particular symbol of $\calXWc$ has occurred with each symbol of $\calXW$ in the pair $[\vxW; \vxWc]$. In particular, for $(a,b)\in \calXW \times \calXWc$, the conditional type is defined as $\typeveccond\left(b | a\right) = \frac{\N{a,b}{\vxW,\vxWc}}{\N{a}{\vxW}}$.

Now, by Chernoff bound, we have
\begin{IEEEeqnarray}{rCl}
\IEEEeqnarraymulticol{3}{l}{\Pr{}_{\mathcal{B}}\left(\Big| \left|\typeveccond\cap\Bin{\vm}\right| - \expect{\left|\typeveccond\cap\Bin{\vm}\right|} \Big| \geq \eps\: \expect{\left|\typeveccond\cap\Bin{\vm}\right|} \right)} \nonumber\\
\label{eq:chernoff-cond-type} 
\qquad & \leq & 2\exp{\left(\frac{-\eps^2\expect{\left|\typeveccond\cap\Bin{\vm}\right|}}{3}\right)}.
\end{IEEEeqnarray}
Note that the expected number of codeword (balls) of a particular conditional type that fall into any message bin is 
\begin{equation}
\label{eq:expect-card-cond-type}
\expect{\left|\typeveccond(b | a)\cap\Bin{\vm}\right|} = \frac{\left|\typeveccond(b | a)\right|}{2^{n\R}},
\end{equation}
for the total number of message bins is equal to the total number of messages. 

Next, to upper bound~\eqref{eq:chernoff-cond-type}, let us find a lower bound on the cardinality of the conditional type class. If we multiply the stochastic matrix associated with the conditional type $\typeveccond$ by the type of $\vxW$, denoted as $\typevec_{\vxW}$, we get the joint type of $\vx = [\vxW;\vxWc]$ denoted as $\typevec_{\vx}$. Now, the conditional type class can be considered as the set $\typeveccond = \left\{\vxWc : \left[\vxW; \vxWc \right] \in \typevec_{\vx} \right\}$. Observe that $\left|\typeveccond\right|$ is constant for a given $\vxW\in\typevec_{\vxW}$, and thus, $\left|\typeveccond\right| = \frac{\left|\typevec_{\vx}\right|}{\left|\typevec_{\vxW}\right|}$. Then, by lower bounding $\left|\typevec_{\vx}\right|$ and upper bounding $\left|\typevec_{\vxW}\right|$ using \cite[Theorem 11.1.3]{CovT:12}, we have
\begin{equation}
\label{eq:card-cond-type-LB}
\frac{1}{(n+1)^{|\calXW||\calXWc|}}\frac{2^{n\Hp{\typevec_{\vx}}}}{2^{n\Hp{\typevec_{\vxW}}}} \leq \left|\typeveccond\right|
\end{equation}
Therefore, if $\R \leq \Hp{\pactscalcond{.}{.}} - \constone$, using~\eqref{eq:expect-card-cond-type} and~\eqref{eq:card-cond-type-LB}, the probability in~\eqref{eq:chernoff-cond-type} can be upper bounded by
\begin{IEEEeqnarray}{rCl}
\IEEEeqnarraymulticol{3}{l}{\Pr{}_{\mathcal{B}}\left(\Big| \left|\typeveccond\cap\Bin{\vm}\right| - \expect{\left|\typeveccond\cap\Bin{\vm}\right|} \Big| \geq \eps \expect{\left|\typeveccond\cap\Bin{\vm}\right|} \right)} \nonumber\\
\label{eq:chernoff-cond-type-bound} 
\qquad & \leq & 2\exp{\left(\frac{-\eps^2{2^{n\left[\Hp{\typevec_{\vx}} - \Hp{\typevec_{\vxW}} - \Hp{\pactscalcond{.}{.}} - \frac{{|\calXW||\calXWc|}\log_2(n+1)}{n} + \constone \right]}}}{3}\right)}.
\end{IEEEeqnarray}
This probability will be close to zero if 
\begin{equation}
\label{eq:chernoff-exponent}
\left[\Hp{\typevec_{\vx}} - \Hp{\typevec_{\vxW}} - \Hp{\pactscalcond{.}{.}} - \frac{{|\calXW||\calXWc|}\log_2(n+1)}{n} + \constone \right] > 0.
\end{equation}

The next task is to characterize $\constone$ such that the aforementioned inequality holds. For that, let $\theta(\typevec_{\vx})$ denote the variation distance between the type $\typevec_{\vx}$ and $\pactscal{.}$. Then, since $\vx$ is strongly typical, it is straightforward to show (using similar arguments as in Lemma 3) that $\theta(\typevec_{\vx}) \leq \frac{\eps}{2}$. Similarly, let $\theta(\typevec_{\vxW})$ denote the variation distance between the type $\typevec_{\vxW}$ and $\pactscalW{.}$. Then, using the strong typicality of $\vxW$, we can show that $\theta(\typevec_{\vxW}) \leq \frac{\eps}{2}$.

Now, observe that
\begin{IEEEeqnarray}{rCl}
\left| \Hp{\typevec_{\vx}} - \Hp{\typevec_{\vxW}} - \Hp{\pactscalcond{.}{.}} \right| & \stackrel{(a)}{=} & 
\left| \Hp{\typevec_{\vx}} - \Hp{\typevec_{\vxW}} - \Hp{\pactscal{.}} + \Hp{\pactscalW{.}} \right| \IEEEeqnarraynumspace \\
& \leq & \left| \Hp{\typevec_{\vx}} - \Hp{\pactscal{.}} \right| + \left| \Hp{\pactscalW{.}} - \Hp{\typevec_{\vxW}} \right| \IEEEeqnarraynumspace \\
& \stackrel{(b)}{\leq} & 2\theta(\typevec_{\vx})\log_2 \frac{\left|\calX\right|}{2\theta(\typevec_{\vx})} + 2\theta(\typevec_{\vxW})\log_2 \frac{\left|\calXW\right|}{2\theta(\typevec_{\vxW})} \\
& \stackrel{(c)}{\leq} & \eps\log_2 \frac{\left|\calX\right|}{\eps} + \eps\log_2 \frac{\left|\calXW\right|}{\eps},
\end{IEEEeqnarray}
where, to obtain (a), we use the fact that the distribution $\pactscal{.}$ can be written as $\pactscal{c} = \pactscalWWc{a,b} = \pactscalcond{b}{a}\pactscalW{a}$ for $c\in\calX$, $(a,b)\in\calXW\times\calXWc$; (b) follows from the result $\left| \Hp{\typevec} - \Hp{\pactscal{.}} \right| \leq 2\theta(\typevec)\log_2 \frac{\left|\calX\right|}{2\theta(\typevec)}$ for $0\leq\theta(\typevec)\leq\frac{1}{4}$ \cite[Lemma 2.7]{CsisK-book:2011}; and (c) follows since $-t\log_2{t}$ is an increasing function of $t$ for $0\leq t\leq \frac{1}{2}$.

Therefore, LHS in equation~\eqref{eq:chernoff-exponent} becomes
\begin{IEEEeqnarray}{rCl}
\IEEEeqnarraymulticol{3}{l}
{\Hp{\typevec_{\vx}} - \Hp{\typevec_{\vxW}} - \Hp{\pactscalcond{.}{.}} - \frac{{|\calXW||\calXWc|}\log_2(n+1)}{n} + \constone} \nonumber\\
\label{eq:chernoff-exponent-bound}
\quad & \geq & 2\eps\log_2{\frac{\eps}{2^{\C}}} - \frac{2^{\C}\log_2(n+1)}{n} + \constone,
\end{IEEEeqnarray}
since $\left|\calXW\right| \left|\calXWc\right| = \left|\calX\right| = 2^{\C}$. Above term~\eqref{eq:chernoff-exponent-bound} is positive if $\constone \geq 2\eps \log_2\left(\frac{2^{\C}}{\eps}\right) + \frac{2^{\C}\log_2(n+1)}{n}$, and in that case, $\left| \typeveccond \cap\:\Bin{\vm}\right| \to \expect{\left|\typeveccond\cap\:\Bin{\vm}\right|} = \frac{\left|\typeveccond\right|}{\left|\vM\right|}$ with high probability.


\section*{Acknowledgment}
Author Swanand Kadhe would like to thank Pak Hou (Howard) Che and Dr. Chung Chan for helpful discussions.

\ifCLASSOPTIONcaptionsoff
  \newpage
\fi



\bibliographystyle{IEEEtran}
\bibliography{IEEEabrv,Nw_deniability_v2}
%
%
%
%

%




\end{document}

%% file: network_deniability_system_42.pdf_tex
\begingroup%
  \makeatletter%
  \providecommand\color[2][]{%
    \errmessage{(Inkscape) Color is used for the text in Inkscape, but the package 'color.sty' is not loaded}%
    \renewcommand\color[2][]{}%
  }%
  \providecommand\transparent[1]{%
    \errmessage{(Inkscape) Transparency is used (non-zero) for the text in Inkscape, but the package 'transparent.sty' is not loaded}%
    \renewcommand\transparent[1]{}%
  }%
  \providecommand\rotatebox[2]{#2}%
  \ifx\svgwidth\undefined%
    \setlength{\unitlength}{334.63515625bp}%
    \ifx\svgscale\undefined%
      \relax%
    \else%
      \setlength{\unitlength}{\unitlength * \real{\svgscale}}%
    \fi%
  \else%
    \setlength{\unitlength}{\svgwidth}%
  \fi%
  \global\let\svgwidth\undefined%
  \global\let\svgscale\undefined%
  \makeatother%
  \begin{picture}(1,0.28793252)%
    \put(0,0){\includegraphics[width=\unitlength]{network_deniability_system_42.pdf}}%
    \put(0.11356118,0.21726787){\color[rgb]{0,0,0}\makebox(0,0)[lb]{\smash{$\vX\sim \pinn{.}$}}}%
    \put(0.25819631,0.20053323){\color[rgb]{0,0,0}\makebox(0,0)[lb]{\smash{$\mathbf{T}=0$}}}%
    \put(0.25819631,0.13837598){\color[rgb]{0,0,0}\makebox(0,0)[lb]{\smash{$\mathbf{T} = 1$}}}%
    \put(0.0466226,0.02601481){\color[rgb]{0,0,0}\makebox(0,0)[lb]{\smash{$M \in \{1,2,\ldots,2^{nR}\}$}}}%
    \put(0.59647517,0.22922119){\color[rgb]{0,0,0}\makebox(0,0)[lb]{\smash{Estimate $\T$ reliably}}}%
    \put(0.60842849,0.10968802){\color[rgb]{0,0,0}\makebox(0,0)[lb]{\smash{Estimate $\T$ better than random}}}%
    \put(0.70405502,0.16945461){\color[rgb]{0,0,0}\makebox(0,0)[lb]{\smash{\textrm{Bob}}}}%
    \put(0.72796166,0.03796812){\color[rgb]{0,0,0}\makebox(0,0)[lb]{\smash{\textrm{Willie}}}}%
    \put(0.13029582,0.16945461){\color[rgb]{0,0,0}\makebox(0,0)[lb]{\smash{Alice}}}%
    \put(0.1183425,0.12164134){\color[rgb]{0,0,0}\makebox(0,0)[lb]{\smash{$\enc$}}}%
    \put(0.42912874,0.27703446){\color[rgb]{0,0,0}\makebox(0,0)[lb]{\smash{$\vX_1$}}}%
    \put(0.42912874,0.24117451){\color[rgb]{0,0,0}\makebox(0,0)[lb]{\smash{$\vX_2$}}}%
    \put(0.42912874,0.20531456){\color[rgb]{0,0,0}\makebox(0,0)[lb]{\smash{$\vX_3$}}}%
    \put(0.57256854,0.02601481){\color[rgb]{0,0,0}\makebox(0,0)[lb]{\smash{$\vXW$}}}%
    \put(0.54866191,0.26508114){\color[rgb]{0,0,0}\makebox(0,0)[lb]{\smash{$\vX = \left[\vX_1^T \: \vX_2^T \: \cdots \: \vX_\C^T \right]^T$}}}%
    \put(0.42912874,0.08578139){\color[rgb]{0,0,0}\makebox(0,0)[lb]{\smash{$\vX_C$}}}%
    \put(0.60842849,0.08578139){\color[rgb]{0,0,0}\makebox(0,0)[lb]{\smash{Given $\T = 1$, gain some information about $\M$
}}}%
    \put(0.59647517,0.20531456){\color[rgb]{0,0,0}\makebox(0,0)[lb]{\smash{Given $\T = 1$, decode $\M$ reliably
}}}%
    \put(0.60842849,0.16945461){\color[rgb]{0,0,0}\makebox(0,0)[lb]{\smash{Decoder}}}%
    \put(0.63233512,0.03796812){\color[rgb]{0,0,0}\makebox(0,0)[lb]{\smash{Estimator}}}%
    \put(-0.00119066,0.15750129){\color[rgb]{0,0,0}\makebox(0,0)[lb]{\smash{Private}}}%
    \put(-0.00119066,0.09773471){\color[rgb]{0,0,0}\makebox(0,0)[lb]{\smash{$r$-bits}}}%
    \put(-0.00119066,0.13359466){\color[rgb]{0,0,0}\makebox(0,0)[lb]{\smash{randomness}}}%
    \put(0.0466226,0.00210817){\color[rgb]{0,0,0}\makebox(0,0)[lb]{\smash{Rate = $\R$ bits / ch use }}}%
    \put(0.42912874,0.02601481){\color[rgb]{0,0,0}\makebox(0,0)[lb]{\smash{$\W\in\bigW$}}}%
  \end{picture}%
\endgroup%

%% file: example_1.pdf_tex
\begingroup%
  \makeatletter%
  \providecommand\color[2][]{%
    \errmessage{(Inkscape) Color is used for the text in Inkscape, but the package 'color.sty' is not loaded}%
    \renewcommand\color[2][]{}%
  }%
  \providecommand\transparent[1]{%
    \errmessage{(Inkscape) Transparency is used (non-zero) for the text in Inkscape, but the package 'transparent.sty' is not loaded}%
    \renewcommand\transparent[1]{}%
  }%
  \providecommand\rotatebox[2]{#2}%
  \ifx\svgwidth\undefined%
    \setlength{\unitlength}{318.62027588bp}%
    \ifx\svgscale\undefined%
      \relax%
    \else%
      \setlength{\unitlength}{\unitlength * \real{\svgscale}}%
    \fi%
  \else%
    \setlength{\unitlength}{\svgwidth}%
  \fi%
  \global\let\svgwidth\undefined%
  \global\let\svgscale\undefined%
  \makeatother%
  \begin{picture}(1,0.36523902)%
    \put(0,0){\includegraphics[width=\unitlength]{example_1.pdf}}%
    \put(0.12861841,0.2962235){\color[rgb]{0,0,0}\makebox(0,0)[lb]{\smash{$\vX\sim \pinn{.}$}}}%
    \put(0.31190869,0.26358276){\color[rgb]{0,0,0}\makebox(0,0)[lb]{\smash{$\mathbf{T}=0$}}}%
    \put(0.31190869,0.19830129){\color[rgb]{0,0,0}\makebox(0,0)[lb]{\smash{$\mathbf{T} = 1$}}}%
    \put(0.15623749,0.0627167){\color[rgb]{0,0,0}\makebox(0,0)[lb]{\smash{$M \in \{1,2,\ldots,2^{n}\}$}}}%
    \put(0.69983127,0.28618019){\color[rgb]{0,0,0}\makebox(0,0)[lb]{\smash{\textrm{Bob}}}}%
    \put(0.69983127,0.12297652){\color[rgb]{0,0,0}\makebox(0,0)[lb]{\smash{\textrm{Willie}}}}%
    \put(0.15623749,0.23345285){\color[rgb]{0,0,0}\makebox(0,0)[lb]{\smash{Alice}}}%
    \put(0.14161675,0.17972901){\color[rgb]{0,0,0}\makebox(0,0)[lb]{\smash{Encoder}}}%
    \put(0.48766649,0.30124515){\color[rgb]{0,0,0}\makebox(0,0)[lb]{\smash{$\vX_1$}}}%
    \put(0.48766649,0.16314973){\color[rgb]{0,0,0}\makebox(0,0)[lb]{\smash{$\vX_2$}}}%
    \put(0.63706063,0.07276){\color[rgb]{0,0,0}\makebox(0,0)[lb]{\smash{$\vec{\mathbf{X}}_W$}}}%
    \put(0.67597842,0.35146166){\color[rgb]{0,0,0}\makebox(0,0)[lb]{\smash{$\vX = \left[\vX_1^T \: \vX_2^T  \right]^T$}}}%
    \put(0.07714648,0.00245688){\color[rgb]{0,0,0}\makebox(0,0)[lb]{\smash{$\enc : \{0,1\}^{n}\times \{0,1\}^{n} \to \{0,1\}^{2 \times n}$}}}%
    \put(0.71291686,0.23239345){\color[rgb]{0,0,0}\makebox(0,0)[lb]{\smash{Decoder}}}%
    \put(0.71121182,0.06858405){\color[rgb]{0,0,0}\makebox(0,0)[lb]{\smash{Estimator}}}%
    \put(-0.00068912,0.20081212){\color[rgb]{0,0,0}\makebox(0,0)[lb]{\smash{Private}}}%
    \put(0.01939749,0.15310643){\color[rgb]{0,0,0}\makebox(0,0)[lb]{\smash{$r=1$-bit}}}%
    \put(-0.0013682,0.18324262){\color[rgb]{0,0,0}\makebox(0,0)[lb]{\smash{randomness}}}%
    \put(0.15783244,0.03500437){\color[rgb]{0,0,0}\makebox(0,0)[lb]{\smash{Rate $R = 1$ bit/ch use
}}}%
    \put(0.44874869,0.07527083){\color[rgb]{0,0,0}\makebox(0,0)[lb]{\smash{$\bigW = \left\{ \{1\},\{2\} \right\}$}}}%
    \put(0.44808857,0.03864065){\color[rgb]{0,0,0}\makebox(0,0)[lb]{\smash{$W \in \bigW$}}}%
  \end{picture}%
\endgroup%

%% file: Random_binning_11.pdf_tex
\begingroup%
  \makeatletter%
  \providecommand\color[2][]{%
    \errmessage{(Inkscape) Color is used for the text in Inkscape, but the package 'color.sty' is not loaded}%
    \renewcommand\color[2][]{}%
  }%
  \providecommand\transparent[1]{%
    \errmessage{(Inkscape) Transparency is used (non-zero) for the text in Inkscape, but the package 'transparent.sty' is not loaded}%
    \renewcommand\transparent[1]{}%
  }%
  \providecommand\rotatebox[2]{#2}%
  \ifx\svgwidth\undefined%
    \setlength{\unitlength}{251.31328125bp}%
    \ifx\svgscale\undefined%
      \relax%
    \else%
      \setlength{\unitlength}{\unitlength * \real{\svgscale}}%
    \fi%
  \else%
    \setlength{\unitlength}{\svgwidth}%
  \fi%
  \global\let\svgwidth\undefined%
  \global\let\svgscale\undefined%
  \makeatother%
  \begin{picture}(1,0.60485116)%
    \put(0,0){\includegraphics[width=\unitlength]{Random_binning_11.pdf}}%
    \put(0.1098293,0.51075794){\color[rgb]{0,0,0}\makebox(0,0)[lb]{\smash{$\typ\left(\vX\right)$}}}%
    \put(0.51092231,0.59033989){\color[rgb]{0,0,0}\makebox(0,0)[lb]{\smash{Message Bins}}}%
    \put(0.73375176,0.51075794){\color[rgb]{0,0,0}\makebox(0,0)[lb]{\smash{$\Bin{1}$}}}%
    \put(0.73375176,0.43117599){\color[rgb]{0,0,0}\makebox(0,0)[lb]{\smash{$\Bin{2}$}}}%
    \put(0.73375176,0.2720121){\color[rgb]{0,0,0}\makebox(0,0)[lb]{\smash{$\Bin{2^{n\Rd}}$}}}%
    \put(0.73375176,0.03326626){\color[rgb]{0,0,0}\makebox(0,0)[lb]{\smash{$\Bin{2^{n\R}}$}}}%
    \put(0.01433097,0.44709238){\color[rgb]{0,0,0}\makebox(0,0)[lb]{\smash{$\typevec_1$}}}%
    \put(-0.00158542,0.38342683){\color[rgb]{0,0,0}\makebox(0,0)[lb]{\smash{$\typevec_2$}}}%
    \put(0.01433097,0.14468099){\color[rgb]{0,0,0}\makebox(0,0)[lb]{\smash{$\typevec_t$}}}%
    \put(0.92474843,0.39934322){\color[rgb]{0,0,0}\makebox(0,0)[lb]{\smash{$\code$}}}%
    \put(0.31355908,0.22426293){\color[rgb]{0,0,0}\makebox(0,0)[lb]{\smash{Random}}}%
    \put(0.31355908,0.19243016){\color[rgb]{0,0,0}\makebox(0,0)[lb]{\smash{Binning}}}%
  \end{picture}%
\endgroup%

%% file: example_2.pdf_tex
\begingroup%
  \makeatletter%
  \providecommand\color[2][]{%
    \errmessage{(Inkscape) Color is used for the text in Inkscape, but the package 'color.sty' is not loaded}%
    \renewcommand\color[2][]{}%
  }%
  \providecommand\transparent[1]{%
    \errmessage{(Inkscape) Transparency is used (non-zero) for the text in Inkscape, but the package 'transparent.sty' is not loaded}%
    \renewcommand\transparent[1]{}%
  }%
  \providecommand\rotatebox[2]{#2}%
  \ifx\svgwidth\undefined%
    \setlength{\unitlength}{338.82965088bp}%
    \ifx\svgscale\undefined%
      \relax%
    \else%
      \setlength{\unitlength}{\unitlength * \real{\svgscale}}%
    \fi%
  \else%
    \setlength{\unitlength}{\svgwidth}%
  \fi%
  \global\let\svgwidth\undefined%
  \global\let\svgscale\undefined%
  \makeatother%
  \begin{picture}(1,0.34345447)%
    \put(0,0){\includegraphics[width=\unitlength]{example_2.pdf}}%
    \put(0.12094701,0.27855535){\color[rgb]{0,0,0}\makebox(0,0)[lb]{\smash{$\vX\sim \pinn{.}$}}}%
    \put(0.293305,0.24786146){\color[rgb]{0,0,0}\makebox(0,0)[lb]{\smash{$\mathbf{T}=0$}}}%
    \put(0.293305,0.18647368){\color[rgb]{0,0,0}\makebox(0,0)[lb]{\smash{$\mathbf{T} = 1$}}}%
    \put(0.14691876,0.05897599){\color[rgb]{0,0,0}\makebox(0,0)[lb]{\smash{$M \in \{1,2,\ldots,2^{n}\}$}}}%
    \put(0.65809008,0.26911108){\color[rgb]{0,0,0}\makebox(0,0)[lb]{\smash{\textrm{Bob}}}}%
    \put(0.65809008,0.11564163){\color[rgb]{0,0,0}\makebox(0,0)[lb]{\smash{\textrm{Willie}}}}%
    \put(0.14691876,0.21952864){\color[rgb]{0,0,0}\makebox(0,0)[lb]{\smash{Alice}}}%
    \put(0.13317006,0.16900914){\color[rgb]{0,0,0}\makebox(0,0)[lb]{\smash{Encoder}}}%
    \put(0.4585798,0.28327749){\color[rgb]{0,0,0}\makebox(0,0)[lb]{\smash{$\vX_1$}}}%
    \put(0.4585798,0.20064009){\color[rgb]{0,0,0}\makebox(0,0)[lb]{\smash{$\vX_2$}}}%
    \put(0.59906337,0.06842026){\color[rgb]{0,0,0}\makebox(0,0)[lb]{\smash{$\vec{\mathbf{X}}_{\W}$}}}%
    \put(0.62621566,0.33049886){\color[rgb]{0,0,0}\makebox(0,0)[lb]{\smash{$\vX = \left[\vX_1^T \: \vX_2^T\: \vX_3^T \right]^T$}}}%
    \put(0.13865502,0.00231034){\color[rgb]{0,0,0}\makebox(0,0)[lb]{\smash{$\enc : \{0,1\}^{n} \to \{0,1\}^{3 \times n}$}}}%
    \put(0.67039519,0.21853242){\color[rgb]{0,0,0}\makebox(0,0)[lb]{\smash{Decoder}}}%
    \put(0.66879184,0.06449338){\color[rgb]{0,0,0}\makebox(0,0)[lb]{\smash{Estimator}}}%
    \put(-0.00064802,0.18883475){\color[rgb]{0,0,0}\makebox(0,0)[lb]{\smash{Private}}}%
    \put(0.01824053,0.14397445){\color[rgb]{0,0,0}\makebox(0,0)[lb]{\smash{$r=0$-bit}}}%
    \put(-0.0012866,0.17231318){\color[rgb]{0,0,0}\makebox(0,0)[lb]{\smash{randomness}}}%
    \put(0.14841858,0.03291654){\color[rgb]{0,0,0}\makebox(0,0)[lb]{\smash{Rate $R = 1$ bit/ch use
}}}%
    \put(0.33226263,0.07078133){\color[rgb]{0,0,0}\makebox(0,0)[lb]{\smash{$\bigW = \left\{ \{1,2\},\{2,3\},\{1,3\} \right\}$}}}%
    \put(0.42136249,0.03633594){\color[rgb]{0,0,0}\makebox(0,0)[lb]{\smash{$\W \in \bigW$}}}%
    \put(0.45739926,0.162863){\color[rgb]{0,0,0}\makebox(0,0)[lb]{\smash{$\vX_3$}}}%
  \end{picture}%
\endgroup%